\documentclass[traditabstract,bibyear]{aa} 
\usepackage{epsfig}
\usepackage{amsmath}
\usepackage{subfigure}
\usepackage{natbib}
\usepackage{multirow}
\usepackage{color}
\usepackage{longtable}
\usepackage{graphicx}
\usepackage{epstopdf}
\usepackage{booktabs}
\usepackage{mhchem}
\usepackage{txfonts}

\newcommand{\jms}{J.~Mol.~Spectr.}

\newcommand{\jmst}{J.~Mol.~Struct.}
\newcommand{\natastro}{Nature Astron.}
\newcommand{\science}{Science}

\newcommand{\kms}{km\,s$^{-1}$}

\begin{document}

\title{The physical and chemical structure of TMC-1. The spatial distribution 
and excitation conditions of carbon, oxygen, and sulfur bearing species: Depletion
of CO and CS}

\author{
J.~Cernicharo\inst{1},
B.~Tercero\inst{2,3},
N.~Marcelino\inst{2,3},
G.~Esplugues\inst{2},
G.~Molpeceres\inst{1}, and
M.~Ag\'undez\inst{1}
}

\institute{Dept. de Astrof\'isica Molecular, Instituto de F\'isica Fundamental (IFF-CSIC),
C/ Serrano 121, 28006 Madrid, Spain. \newline \email jose.cernicharo@csic.es
\and Observatorio Astron\'omico Nacional (IGN), C/ Alfonso XII, 3, 28014, Madrid, Spain.
\and Centro de Desarrollos Tecnol\'ogicos, Observatorio de Yebes (IGN), Cerro de la Palera s/n, 19141 Yebes, Guadalajara, Spain.
}

\date{Received 23/06/2026; accepted 14/10/2026}

\abstract{We present a detailed study of the physical and chemical structure of TMC-1 from the observed spatial distribution of HC$_3$N, NNH$^+$, CN, and their isotopologues, the O-bearing species C$^{18}$O, C$^{17}$O, CH$_3$OH, CH$_3$CHO, and H$_2$CO, and the S-bearing species CS, HCS$^+$, H$_2$CS, and H$_2$CCS. The well known filamentary structure of TMC-1 can be interpreted as a single cloud with a velocity gradient perpendicular to the filament, or less likely as due to the alignment of three independent clouds in the line sight with velocities of 5.3-5.7, 5.7-5.9, and 5.9-6.3 \kms. The main contribution to the emission of all studied C-bearing species is found at 5.83 \kms. We conclude that the three velocity components are dynamically connected and that C-bearing molecules show a different behavior of their abundances compared to O-bearing species, which have emission peaks at 5.7 and 6.25 \kms\, and appear at different spatial locations than cyanopolyynes and C$_n$H and C$_n$N radicals. The detection of $^{13}$C$^{18}$O, with exactly the same line profile than that of C$^{18}$O and C$^{17}$O, excludes the possibility that the decrease of their emission at 5.83 \kms\, could be due to self absorption or absorption by a foreground cloud component. Hence, as indicated by the spatial distribution of C$^{18}$O and C$^{17}$O, CO is significantly depleted at the velocities and positions where the emission of the carbon chains and radicals is found. This suggests that the emission of O-bearing species is coming  from the external layers of the carbon-rich filaments. A comparison between S-bearing species and other molecules indicates that CS, CCS, HCS$^+$, and H$_2$CS are significantly depleted in the zone of CO depletion. A clear increase of the deuteration of several molecular species is found in the zones where CO and CS are depleted. Radiative trapping effects modify the emission distribution of abundant species with a high dipole moment and, hence, the determination of their abundances requires the observation of minor isotopologues. Our chemical models reveal that CO and CS depletion can be explained by adsorption onto grains, whereas such a process on carbon chains is slower than gas-phase conversion explaining the segregation found by observations.
Finally, the $^{12}$C/$^{13}$C abundance ratio derived from CO, CH$_3$OH, and CN is 65$\pm$7, 45$\pm$5, and 72$\pm$13, respectively.}


\titlerunning{Physical and Chemical Structure of TMC-1}
\authorrunning{Cernicharo et al.}

\maketitle

\section{Introduction}
TMC-1 is a quiescent dense filament in the Taurus molecular cloud complex. It lies inside the Heiles' Cloud 2 \citep[][hereafter referred as HCL2]{Heiles1968} at a distance of  $\sim$140 pc 
\citep{Kirk2024}. HCL2 covers more than 1 square degree in the sky and has a 
$\gamma$-shape with a double peaked distribution of velocities at 5.35 and 6.25 \kms. TMC-1 has a filamentary structure \citep[][and references therein]{Lique2006} and is characterized by 
a very rich molecular spectrum with strong chemical differences along its ridge. Consequently, it is considered as an ideal source to study the chemistry of dark clouds 
\citep[see, e.g.,][]{Pratap1997,Fosse2001,Loison2017}. The position observed in this work, the cyanopolyyne peak (hereafter referred as CP), is particularly rich in carbon chains, 
including radicals (C$_n$H, C$_n$N, and their anions) and cyanopolyynes (HC$_{2n+1}$N, and their protonated species). Although several cloudlets in Taurus, and also in particular 
in HCL2, show a rich carbon chemistry \citep[see, e.g.,][]{Cernicharo1984a,Agundez2023a}, TMC-1 stands out as the source with the largest column densities of carbon chains.

TMC-1 has been previously surveyed in the 3\,mm band with the IRAM\,30m telescope \citep[see][]{Marcelino2007,Marcelino2009,Cernicharo2012b}. It was also observed with the 
Nobeyama 45m telescope at frequencies below 50 GHz \citep{Kaifu2004}. Recently the source was also observed in these bands with the Green Bank 100m telescope \citep[][the GOTHAM 
line survey]{McGuire2018,McGuire2021}. The ultra sensitive line survey QUIJOTE\footnote{\textbf{Q}-band \textbf{U}ltrasensitive 
\textbf{I}nspection \textbf{J}ourney to the \textbf{O}bscure \textbf{T}MC-1 \textbf{E}nvironment} \citep{Cernicharo2021a} has been obtained using the Q-band broad band receivers 
and spectrometers installed at the 40m telescope of the Yebes Observatory. The QUIJOTE line survey has provided in the last six years the robust and unambiguous detection of 
more than 70 new molecules in TMC-1. 

With our deep survey observations at 3 and 7\,mm we have found that the chemical conditions in TMC-1 permit the presence, with very large abundances, of saturated hydrocarbon species such as CH$_3$CHCH$_2$ \citep{Marcelino2007}, CH$_3$CCH \citep[see, e.g.,][and references therein]{Cabezas2021a,Agundez2021a,Agundez2023b}, and long pure hydrocarbons such as CH$_2$CCH \citep{Agundez2021b,Agundez2022a}, CH$_2$CHCCH \citep{Cernicharo2021b}, CH$_2$CCHC$_4$H and HCCCHCCC \citep{Fuentetaja2022a,Fuentetaja2022b}, $c$-C$_5$H \citep{Cabezas2022a}, $c$-C$_6$H$_4$ \citep{Cernicharo2021a}, cyclopentadiene 
\citep[$c$-C$_5$H$_6$;][]{Cernicharo2021c}, and indene \citep[$c$-C$_9$H$_8$;][]{Cernicharo2021c,Burkhardt2021}. Recently, two new PAHs, phelanene \citep[$c$-C$_{13}$H$_{10}$,][]{Cabezas2025a} and 1$H$-cyclopen[cd]indene \citep[$c$-C$_{11}$H$_8$,][]{Fuentetaja2026}, have been found with the QUIJOTE line survey. In addition, the presence of benzene ($c$-C$_6$H$_6$) and naphthalene ($c$-C$_{10}$H$_8$) has been inferred from the detection of their CN and CCH derivatives \citep{McGuire2018,McGuire2021,Loru2023,Cernicharo2024a}. The CN functionalized forms of indene have also been claimed in the same source by \citet{Sita2022}, but have not been confirmed yet with QUIJOTE. Recently, the cyano derivatives of acenaphthylene, pyrene, and coronene have been found in TMC-1 \citep{Cernicharo2024a,Cernicharo2026a,Wenzel2024,Wenzel2025a,Wenzel2025b}. 

TMC-1 is also a source where a significant number of S-bearing species have been detected. In particular, CS, CCS and CCCS are 
among the most abundant molecules in cold dark clouds \citep{Saito1987,Yamamoto1987,Kaifu1987}. These species are also observed in the circumstellar envelopes of carbon-rich evolved stars \citep{Cernicharo1987b}. The abundance and number of sulfur-bearing species in cold clouds is limited, to a large extent, by the significant depletion of sulfur in the cloud \citep[][and references therein]{Fuente2019,Navarro2020, Fuente2023}. In recent years, the models for the chemical composition of TMC-1 have been progressively refined with the discovery of new molecules with the QUIJOTE line survey such as C$_4$S, C$_5$S (previously detected only toward evolved stars), HCCS, H$_2$CCS, H$_2$CCCS, HC$_4$S, NCS, HCSCN, and HCSCCH \citep{Cernicharo2021d,Cernicharo2021e,Fuentetaja2022b}, and cations such as HCCS$^+$ \citep{Cabezas2022b} and HCCCS$^+$ \citep{Cernicharo2021f}. A detailed study of several of these S-bearing species, including their spatial distribution, is provided in \citet{Fuentetaja2025}.

The QUIJOTE line survey has reached an unprecedented level of sensitivity that permits to detect the isotopologues of all molecular species with intensity lines larger than 50 mK, which applies to many polyatomic species \citep{Cernicharo2024b,Tercero2024}. The study of the $^{13}$C, D, $^{34}$S and $^{15}$N isotopologues of species such as CCS, CCCS, C$_4$H, HC$_3$N, HNCCC, HCCNC, HC$_5$N, CH$_2$CN, CH$_3$C$_3$N and CH$_3$C$_4$H has been the subject of detailed studies in the last years using QUIJOTE data \citep{Cernicharo2020a,Cernicharo2024b,Cabezas2021a,Cabezas2021b,Cabezas2022c,Cabezas2025b,Fuentetaja2025,Tercero2024}. 
The rotational lines of the isotopologues of some species are strong enough to derive their spatial distribution using the SANCHO\footnote{\textbf{S}urveying the \textbf{A}rea of the \textbf{N}eighbour TMC-1 \textbf{C}loud through \textbf{H}eterodyne \textbf{O}bservations} spatial maps \citep{Cernicharo2023}, which are absolutely necessary to interpret the data in terms of the physical and velocity structure of the cloud and to derive precise molecular abundances (see Sect.\,\ref{obs_maps}). As an example of the sensitivity of the SANCHO maps it is worth noting that the emission of benzonitrile has been mapped successfully in spite of the weakness of its lines \citep{Cernicharo2023}. The spatial distribution of integrated velocity intensity of HC$_3$N and its isotopologues has been analyzed with these data by \citet{Tercero2024} and, as previously commented, the spatial extent of different S-bearing species has been analysed by \citet{Fuentetaja2025}.

This paper belongs to a series of works related to the abundance and spatial distribution of several abundant molecular species and their isotopologues in TMC-1. This work deals mainly with the physical and chemical structure of the cloud and focus on the depletion of CO and CS and their impact on the chemistry. The observations are described in Sect.\,\ref{sec:Observations} and the line identification methods are delineated in Sect.\,\ref{app:line_identification}. The velocity structure of the cloud is analysed in Sect.\,\ref{sec:Ancilliary}. Sect.\,\ref{Maps} is devoted to a comparison of the spatial distribution of carbon chains and O- and S-bearing species, the study of CO and CS depletion
and its subsequent impact on their chemistry, and to the analysis of the spatial distribution of some COMs. The chemical modeling of the cloud is discussed in Sect.\,\ref{sec:Discussion}.

\section{Observations}\label{sec:Observations}

The observational data used in this work are part of QUIJOTE \citep{Cernicharo2021a}, a spectral line survey of TMC-1 in the Q band carried out with the 
Yebes\,40m telescope\footnote{\texttt{https://rt40m.oan.es/rt40m\_en.php}}
at the position 
$\alpha_{J2000}=4^{\rm h} 41^{\rm  m} 41.9^{\rm s}$ and $\delta_{J2000}=+25^\circ 41' 27.0''$, corresponding to the cyanopolyyne peak in TMC-1 (for references concerning line observations and spatial maps of TMC-1 before J2000, this position corresponds to $\alpha_{J1950}=4^{\rm h} 38^{\rm  m} 38^{\rm s}$ and $\delta_{J1950}=
+25^\circ 35' 45.0''$, see, e.g., \citealt{Cernicharo1987a}). The receiver was built within the Nanocosmos project\footnote{\texttt{https://nanocosmos.iff.csic.es/}} and consists of two cold high-electron mobility transistor amplifiers covering the 31.0-50.3 GHz band with horizontal and vertical polarizations. Receiver temperatures achieved in the 2019 and 2020 runs vary from 22 K at 32 GHz to 42 K at 50 GHz. Some power adaptation in the down-conversion chains performed in 2021 reduced the receiver temperatures to 16\,K at 32 GHz and 30\,K at 50 GHz. The backends are $2\times8\times2.5$ GHz fast Fourier transform spectrometers with a spectral resolution of 38 kHz, providing the whole coverage of the Q-band in both polarizations. The velocity resolution varies between 
0.36 \kms at 31 GHz, and 0.22\kms at 50 GHz. A more detailed description of the system is given in \citet{Tercero2021}. 

The data of the QUIJOTE line survey presented here were gathered in several observing runs between November 2019 and July 2024.  
All observations were performed using frequency-switching observing mode with a frequency throw of 8 and 10 MHz. The total observing time on the source for data taken with frequency throws of 8 MHz and 10 MHz is 736.6 and 772.6 hours, respectively. Hence, the total observing time on source is 1509.2 hours. The measured sensitivity varies between 0.08 mK at 31 GHz and 0.2 mK at 49.5 GHz \citep{Agundez2026}. This sensitivity is around 50 times better than that of previous line surveys in the Q band of TMC-1 \citep{Kaifu2004}. For each frequency throw, different local oscillator frequencies were used in order to remove possible side band effects in the down conversion chain. A detailed description of the QUIJOTE line survey is provided in \citet{Cernicharo2021a}. The data analysis procedure has been described by \citet{Cernicharo2022a}.

The main beam efficiency measured during our observations varies from 0.66 at 32.4 GHz to 0.50 at 48.4 GHz \citep{Tercero2021} and can be given across the Q band by $B_{\rm eff}$\,=\,0.797 exp[$-$($\nu$(GHz)/71.1)$^2$]. The forward telescope efficiency has varied from 0.93 to 0.97. The telescope beam size at half power intensity (hereafter referred as HPBW) of 36.4$''$ at 48.4 GHz and of 54.4$''$ at 32.4 GHz. 
At the distance of TMC-1 of 140 pc, the spatial resolution of the Yebes 40m telescope is of 
(8-11)$\times$10$^{16}$ cm (i.e., 0.026-0.036 pc).

Data in the millimeter domain for the species studied in this work have been taken with the IRAM\,30m 
telescope\footnote{\texttt{https://iram-institute.org/observatories/30-\\meter-telescope/}}. 
These data correspond to a 3\,mm line 
survey that covers the full available band at the telescope, between 71.6 GHz and 117.6 GHz. The EMIR E0 receiver was connected to the Fourier transform 
spectrometers in its narrow mode, which provide a spectral resolution of 49 kHz and a total bandwidth of 7.2 GHz per spectral setup. 
The corresponding velocity resolution varies between 0.22 at 72 GHz and 0.12 at 115 GHz.
Observations were performed in several runs. Prior to 2012, the old A, B, C receivers were used to cover some selected frequency ranges \citep{Marcelino2007,Marcelino2009,Cernicharo2012b} but these data have limited sensitivity. Between January and May 2012, we completed the scan 82.5-117.6 GHz \citep{Cernicharo2012b}. In August 2018, after the upgrade of the E090 receiver, we extended the survey down to 71.6 GHz. More recent high sensitivity observations in 2021 have been used to improve the signal to noise ratio in several frequency windows \citep{Cernicharo2021g,Agundez2022a,Agundez2022b,Cabezas2022b}. The final 3\,mm line survey has a sensitivity of 2-10 mK, but can be as low as 0.6 mK at some selected frequencies ranges. All the observations were performed using the frequency switching method with a frequency throw of 7.14 MHz. The IRAM\,30m beam varies between 21$''$ and 34$''$ at 117 and 72 GHz, respectively, while the beam efficiency takes values of 
0.78 and 0.83 at the same frequencies, following the relation B$_{\rm eff}$\,=\,0.871 exp[$-$($\nu$(GHz)/359)$^2$]. The forward efficiency 
at 3\,mm is 0.95. At the distance of TMC-1, the spatial resolution provided by the IRAM 30m
telescope is 4 and 11$\times$10$^{16}$ cm at 115 and 72 GHz, respectively (i.e., 0.013-0.036 pc).

The intensity scale utilized in this study is the antenna temperature corrected for atmospheric and telescope losses, $T_A^*$. Calibration in both telescopes 
was performed using two absorbers at different temperatures and the atmospheric transmission model ATM \citep{Cernicharo1985,Pardo2001,Pardo2025}. The absolute 
calibration uncertainty is 10\,$\%$. However, the relative calibration between lines within the QUIJOTE survey is certainly better as all them are observed 
simultaneously. All the spectral data were analysed with the GILDAS package\footnote{\texttt{http://www.iram.fr/IRAMFR/GILDAS}}.

The observed lines are indicated in Table \ref{tab:mapped_lines}, which provides the frequency and the angular and velocity resolution for each of them.

\begin{figure*}
\centering
\includegraphics[width=0.95\textwidth]{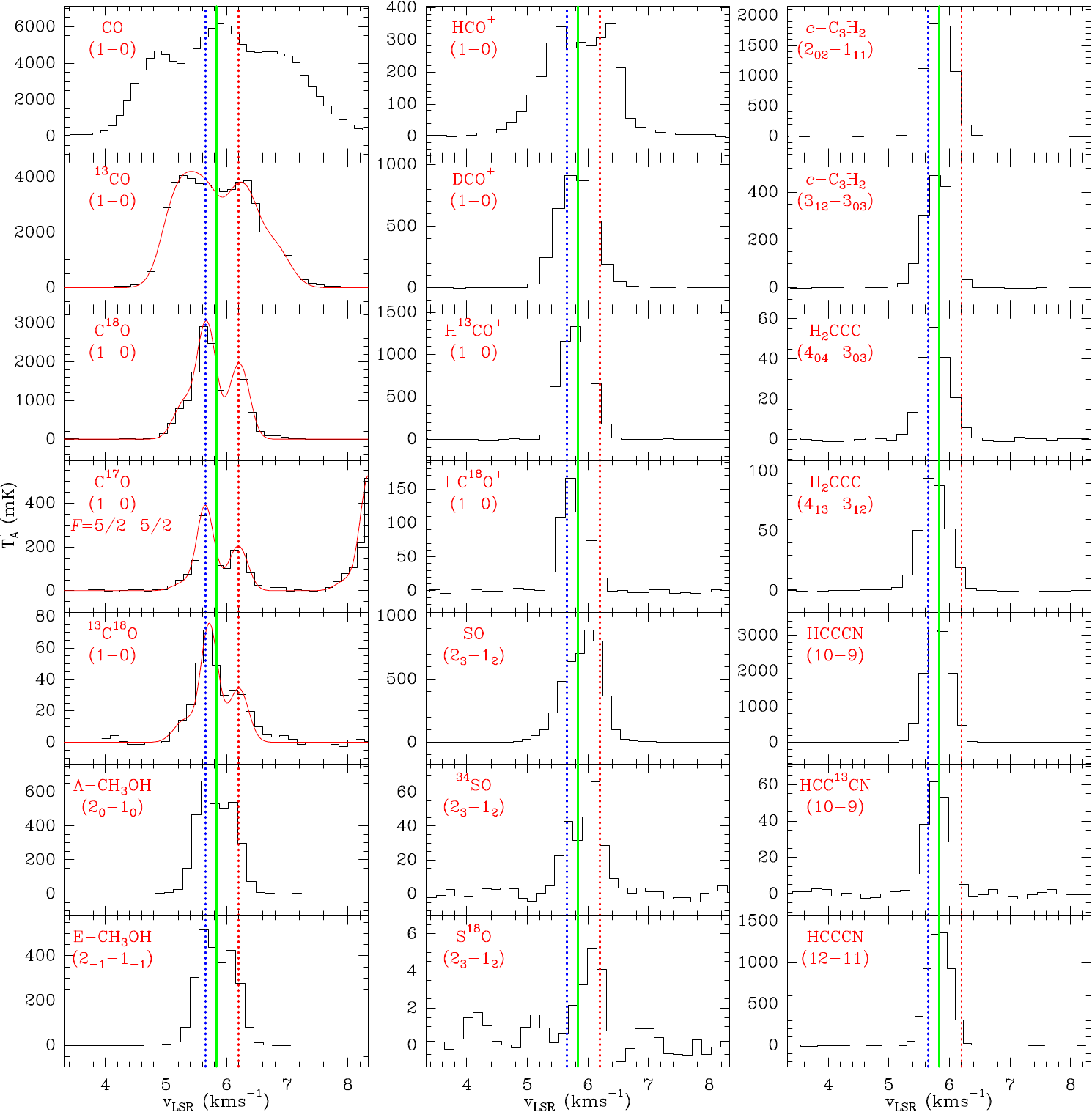}
\caption{Line profiles of assorted molecules (see top-left corner of each panel) observed with the IRAM\,30m telescope at the $CP$ 
position of TMC-1. The abscissa corresponds to the Local Standard of Rest velocity ($V_{\rm LSR}$). Blue, green, and red vertical lines 
correspond to velocities of 5.7, 5.83, and 6.25 km\,s$^{-1}$, respectively. The ordinate is the antenna temperature corrected for 
atmospheric and telescope losses in mK. The line profile obtained from an LTE model with $T_k$\,=\,9 K for the $J$\,=\,1-0 line of 
$^{13}$CO, C$^{18}$O, C$^{17}$O, and $^{13}$C$^{18}$O 
is shown as a red line in the corresponding panel. The physical parameters used are given in Table\,\ref{CO_columndensities} 
(see Sect.\,\ref{sec:CO_line_profiles}). The $J$\,=\,4-3 C$_3$N$^-$ line is broadened due to the hyperfine structure.}
\label{fig:ANC_SPEC}
\end{figure*}

\begin{table*}
\centering
\tiny
\caption{Summary of observed lines depicted and/or mapped in this work}\label{tab:mapped_lines}
\label{table:obs_summary}
\begin{tabular}{llccccccccc}
\hline\hline

Molecule & Transiton & Frequency &  \multicolumn{2}{c}{Yebes 40m} & \multicolumn{2}{c}{IRAM 30m}  & HPBW  & vel res.  & Figure & References \\
         &           &  ( MHz)   & single-point   &   map     &   single-point   &   map       & ('')  & (km\,s$^{-1}$) & number &  \\

\hline

$^{12}$CO           &  $1-0$                   &  115271.202  &   N     &    N     &   Y      &   N       &  21.3  &   0.127  & 1           &  \\
$^{13}$CO           &  $1-0$                   &  110201.354  &   N     &    N     &   Y      &   N$^a$   &  22.3  &   0.132  & 1           &  \\
C$^{18}$O           &  $1-0$                   &  109782.176  &   N     &    N     &   Y      &   Y       &  22.4  &   0.133  & 1, 2       &  \\
C$^{17}$O           &  $1-0$                   &  112359.285  &   N     &    N     &   Y      &   Y       &  21.9  &   0.130  & 1, 2       &  \\
$^{13}$C${18}$O     &  $1-0$                   &  104711.396  &   N     &    N     &   Y      &   N$^a$   &  23.5  &   0.140  & 1           &  \\
HCO$^+$             &  $1-0$                   &  89188.526   &   N     &    N     &   Y      &   N$^a$   &  27.6  &   0.164  & 1           & R1 \\
DCO$^+$             &  $1-0$                   &  72039.312   &   N     &    N     &   Y      &   N$^a$   &  34.1  &   0.203  & 1           & R1 \\
H$^{13}$CO$^+$      &  $1-0$                   &  86754.288   &   N     &    N     &   Y      &   N$^a$   &  28.4  &   0.169  & 1           & R1 \\
HC$^{18}$O$^+$      &  $1-0$                   &  85162.228   &   N     &    N     &   Y      &   N$^a$   &  28.9  &   0.172  & 1           & R1 \\
SO                  & $2_3-1_2$                &  99299.890   &   N     &    N     &   Y      &   N       &  24.8  &   0.147  & 1           & \\
$^{34}$SO           & $2_3-1_2$                &  97715.404   &   N     &    N     &   Y      &   N$^a$   &  25.2  &   0.150  & 1           & \\
S$^{18}$O           & $2_3-1_2$                &  93267.384   &   N     &    N     &   Y      &   N$^a$   &  26.4  &   0.157  & 1           & \\
c-C$_3$H$_2$ (para) & $2_{0,2}-1_{1,1}$        &  82093.548   &   N     &    N     &   Y      &   N$^a$   &  30.0  &   0.178  & 1           & R2\\
c-C$_3$H$_2$ (ortho)& $3_{1,2}-3_{0,3}$        &  82966.196   &   N     &    N     &   Y      &   N       &  29.7  &   0.176  & 1           & R2\\
H$_2$CCC (para)     & $4_{0,4}-3_{0,3}$        &  83165.345   &   N     &    N     &   Y      &   N       &  29.6  &   0.176  & 1           & R2\\
H$_2$CCC (ortho)    & $4_{1,3}-3_{1,2}$        &  83933.699   &   N     &    N     &   Y      &   N       &  29.3  &   0.174  & 1           & R2\\
HCCCN               &  $5-4$                   &  45490.311   &   N$^a$ &    Y     &   N      &   N       &  38.7  &   0.250  & 2           & R3\\
HCCCN               &  $9-8$                   &  81881.462   &   N     &    N     &   N$^a$  &   Y       &  30.0  &   0.179  & 2, 4       & R3\\
HCCCN               &  $10-9$                  &  90978.989   &   N     &    N     &   Y      &   N       &  27.0  &   0.161  & 1           & R3\\
HCCCN               &  $12-11$                 &  109173.637  &   N     &    N     &   Y      &   N$^a$   &  22.5  &   0.135  & 1           & R3\\
HCC$^{13}$CN        &  $5-4$                   &  45301.707   &   N$^a$ &    Y     &   N      &   N       &  38.9  &   0.251  & 3           & R3\\
HCC$^{13}$CN        &  $10-9$                  &  90601.789   &   N     &    N     &   Y      &   N       &  27.2  &   0.162  & 1           & R3\\
DCCCN               &  $5-4$                   &  42215.582   &   N$^a$ &    Y     &   N      &   N       &  41.7  &   0.270  & 3           & R3\\
CH$_3$OH (E)        & $4_{-1}-3_{0}$           &  36169.261   &   Y     &    N$^a$ &   N      &   N       &  48.7  &   0.315  & B.1         & \\
CH$_3$OH (A)        & $1_{0,1}-0_{0,1}$        &  48372.460   &   Y     &    N$^a$ &   N      &   N       &  36.4  &   0.236  & B.1         & \\
CH$_3$OH (E)        & $1_{0}-0_{0}$            &  48376.887   &   Y     &    N$^a$ &   N      &   N       &  36.4  &   0.235  & B.1         & \\
CH$_3$OH (E)        & $2_{-1}-1_{-1}$          &  96739.358   &   N     &    N     &   Y      &   N$^a$   &  25.4  &   0.152  & 1, B.1     & \\
CH$_3$OH (A)        & $2_{0,2}-1_{0,1}$        &  96741.371   &   N     &    N     &   Y      &   Y       &  25.4  &   0.152  & 1, 2, B.1 & \\
CH$_3$OH (E)        & $2_{0}-1_{0}$            &  96744.545   &   N     &    N     &   Y      &   N$^a$   &  25.4  &   0.152  & B.1         & \\
CH$_3$OH (E)        & $2_{1}-1_{1}$            &  96755.501   &   N     &    N     &   Y      &   N$^a$   &  25.4  &   0.152  & B.1         & \\
CH$_3$OH (E)        & $0_{0}-1_{-1}$           & 108893.945   &   N     &    N     &   Y      &   N$^a$   &  22.6  &   0.135  & B.1         & \\
$^{13}$CH$_3$OH (E) & $2_{-1}-1_{-1}$          &  94405.163   &   N     &    N     &   Y      &   N$^a$   &  26.1  &   0.156  & B.1         & \\
$^{13}$CH$_3$OH (A) & $2_{0,2}-1_{0,1}$        &  94407.129   &   N     &    N     &   Y      &   N$^a$   &  26.1  &   0.156  & B.1         & \\
CS                  & $2-1$                    &  97980.953   &   N     &    N     &   N$^a$  &   Y       &  25.1  &   0.150  & 3           & R4 \\
C$^{34}$S           & $2-1$                    &  96412.952   &   N     &    N     &   N$^a$  &   Y       &  25.5  &   0.152  & 3           & R4 \\
$^{13}$CS           & $2-1$                    &  92494.271   &   N     &    N     &   N$^a$  &   Y       &  26.6  &   0.160  & 3           & R4 \\
CN                  & $N=1-0$                  &  113490.970  &   N     &    N     &   Y      &   N$^a$   &  21.7  &   0.129  & D1, D2    & \\
$^{13}$CN           & $N=1-0$                  &  108780.201  &   N     &    N     &   Y      &   N$^a$   &  22.6  &   0.135  & D1         & \\
C$^{15}$N           & $N=1-0$                  &  110024.590  &   N     &    N     &   Y      &   N$^a$   &  22.4  &   0.134  & D1         & \\
HCS$^+$             & $2-1$                    &  85347.875   &   N     &    N     &   N$^a$  &   Y       &  28.8  &   0.172  & 3           & R4 \\
H$_2$CS  (ortho)    & $3_{1,3}-2_{1,2}$        &  101477.805  &   N     &    N     &   N$^a$  &   Y       &  24.2  &   0.145  & 3           & R4 \\
CH$_3$CHO (A)       & $2_{0,2}-1_{0,1}$        &  38512.079   &   N$^a$ &    Y     &   N      &   N       &  45.8  &   0.296  & 2           & \\
H$_2$CO (para)      & $1_{0,1}-0_{0,0}$        &  72837.949   &   N     &    N     &   N$^a$  &   Y       &  33.8  &   0.202  & 2           & \\
HNCO                & $4_{0,4}-3_{0,3}$        &  87925.237   &   N     &    N     &   N$^a$  &   Y       &  28.0  &   0.167  & 5           & \\
H$_2$CCO (ortho)    & $4_{1,3}-3_{1,2}$        &  81586.239   &   N     &    N     &   N$^a$  &   Y       &  30.2  &   0.180  & 5           & \\
CH$_3$CO$^+$        & $2_0-1_0$                &  36537.761   &   N$^a$ &    Y     &   N      &   N       &  48.2  &   0.312  & 5           & \\
CCCO                & $5-4$                    &  48108.474   &   N$^a$ &    Y     &   N      &   N       &  36.6  &   0.237  & 5           & \\
HCCCO$^+$           & $5-4$                    &  44605.633   &   N$^a$ &    Y     &   N      &   N       &  39.5  &   0.255  & 5           & \\
HCO                 & $1_{0,1}-0_{0,0}$        &  86805.780   &   N     &    N     &   N$^a$  &   Y       &  28.3  &   0.169  & 5           & \\
HOCO$^+$            & $2_{0,2}-1_{0,1}$        &  42766.189   &   N$^a$ &    Y     &   N      &   N       &  41.2  &   0.266  & 5           & \\
NNH$^+$             & $J=1-0$, $F_{\rm 1}=0-1$ &  93176.261   &   N     &    N     &   N$^a$  &   Y       &  26.4  &   0-158  & 3           & R5\\
\hline
\hline
\end{tabular}
\tablefoot{
\tablefoottext{a}{The line is not shown in this paper but it has been already observed as reported in the references.}
R1: \citet{Marcelino2026};
R2: Tercero \& Cernicharo, in prep;
R3: \citet{Tercero2024}; 
R4: \citet{Fuentetaja2025};
R5: \citet{Cernicharo2026b}. 
}
\end{table*}
\normalsize

\subsection{Maps}\label{obs_maps}

The modeling of the observed emission in a frequency sweep is often tackled with a very limited information on its spatial extent. 
In the QUIJOTE line survey, as in any other line survey, the only available spatial information is provided by the variation of the 
HPBW of the telescope with the frequency across the line survey. The spatial sizes of the observed molecules, together with the 
issues related to the line opacities and radiative transfer, can only be addressed through spatial mapping of the molecular emission. 
To overcome these issues, the QUIJOTE line survey is being complemented with high sensitivity maps obtained with the Yebes\,40m and 
IRAM\,30m radio telescopes. At the 40m telescope $\sim$100 hours of observing time have been devoted to cover a region of 
320$''$\,$\times$\,320$''$ around the $CP$  position of TMC-1. The maps are fully sampled and cover the whole Q band, as in QUIJOTE. 
We call these supplementary spatial data SANCHO \citep{Cernicharo2023}, and represent a faithful companion to the QUIJOTE line survey. 
The final goal of these maps is to permit the study of the spatial distribution of any QUIJOTE's line having an intensity $\geq$\,20 mK 
with a signal to noise $\geq$\,10, which means all lines of abundant species together with their $^{13}$C, $^{34}$S, D, and $^{15}$N 
isotopologues. The present sensitivity of SANCHO along the Q band in the central position is $\sim$\,3 mK (see Table A.1 of 
\citealt{Cernicharo2023}), which permits to obtain the spatial distribution of several of the new molecules discovered with 
QUIJOTE, including cations, anions, radicals and sulfur-bearing species. The selected final sampling of the maps is 20$''$ in right 
ascension and declination, while the tolerance adopted in adding all the on-the-fly data was 40$''$, i.e., the averaged size of the 
telescope beam in the Q band \citep{Cernicharo2023}. The velocity resolution of the maps is 0.25 \kms\, at 45 GHz and of 0.37 \kms\, 
at 31 GHz. 

Maps for HC$_3$N and its isotopologues have been shown by \citet{Tercero2024} and the spatial distribution of several 
S-bearing species has been presented by \citet{Fuentetaja2025}. Maps of C$_6$H$^-$, C$_6$H, NNH$^+$, and HC$_3$N
have been presented by \citet{Cernicharo2026b}.
Maps of C$_6$H$_5$CN, HC$_5$N, HC$_7$N, C$_4$H, C$_6$H, C$_3$N, CH$_3$CN, HCCNC, HNCCC, $c$-C$_3$H$_2$, and CH$_3$CCH have been shown by \citet{Cernicharo2023}.
A detailed study of the spatial emission of several lines of cyclic and linear C$_3$H$_2$, together with those of their isotopologues, will be published in a forthcoming work (Tercero \& Cernicharo, in prep).

The maps at 3\,mm taken with the IRAM\,30m radio telescope cover the same spatial extent than those gathered with 
the Yebes\,40m telescope. However, they are not as sensitive. Nevertheless, these maps at 3\,mm permit to map the 
emission of the isotopologues of the most abundant molecules. Around 70 hours of observing time have been used in 
three different frequency settings with the IRAM\,30m telescope. The sampling of the data, and the tolerance in 
adding on-the-fly data, is 12$''$. Detailed description of the observations and procedures for the analysis of the 
IRAM\,30m maps have been provided by \citet{Marcelino2026} who analyzed maps of HCO$^+$, HCN and their isotopologues.
The spectral resolution of the maps taken with the IRAM\,30m radio telescope is 48 kHz, which corresponds to a velocity 
resolution of 0.13 \kms\, and 0.18 \kms\, at 110 GHz and 80 GHz, respectively. The sensitivity of these maps depends 
on the frequency and varies between 10 and 20 mK.

The maps that are shown in  this paper cover frequencies from 31 to 115 GHz. Hence, the velocity resolution has a
significant variation depending on the frequency of the mapped molecular line. The same applies to the angular resolution
for each map (see above for the adopted spatial sampling). In order to have a coherent view of the molecular emission at different velocities we have 
adopted for all maps a starting velocity of 5.3 \kms and a velocity step of 0.2 \kms (up to 6.3 \kms). While this choice
is well adapted for frequencies above 50 GHz, at low frequencies some correlation might 
appear between adjacent velocities in the maps. Nevertheless, HC$_3$N has been mapped in its $J$=5-4 (45.5 GHz) and $J$=9-8 (81.9 GHz)
rotational transitions (see Fig. \ref{fig:ANC_maps}). No significant correlation is seen in the velocity maps of the  $J$=5-4 line.

The selected mapped molecular lines are summarized in Table \ref{tab:mapped_lines}. They are
shown in Fig. \ref{fig:ANC_maps}, \ref{fig:ANC_maps_2}, \ref{fig:ANC_maps_3}, and \ref{fig:cn_vel}.

\section{Line profiles} \label{sec:Ancilliary}
Line identification and line modelling are discussed in detail in Appendix \ref{app:line_identification}. 
When frequencies
and collisional rates are available for a molecule then it is possible to predict its line intensities
under the Large Velocity Gradient (LVG) approximation using the formalism described by \citet{Goldreich1974}. When these rates are not available, 
MADEX can adopt Local Thermodynamic Equilibrium (LTE) to compute line intensities, or it
can use generic collisional rates computed adopting very simple assumptions.
Although the density in TMC-1 is not
very high \citep{Agundez2023b,Fuentetaja2025,Marcelino2026}, species with low dipole moment are very close to thermalisation at the kinetic temperature of the cloud of 9\,K \citep{Agundez2023b}. This is the case of the lines of CO discussed below, and that of the several hydrocarbons and PAHs found with 
the QUIJOTE survey \citep{Cernicharo2021a,Cernicharo2021b,Cernicharo2021c}. The method adopted for the analysis of each molecule is discussed when deriving column densities for them (see Appendices \ref{app:line_identification},
\ref{app:line_profiles}, \ref{app:methanol}, \ref{app:two_layer}, and \ref{low_opacity_CN}).

\begin{table*}
\caption{Derived column densities of the isotopologues of CO toward the $CP$  position of TMC-1}
\label{CO_columndensities}
\centering
\small
\begin{tabular}{{cccccccc}}
\hline
$v_{\rm LSR}$$^a$  & $\Delta$v$^b$ & N($^{13}$CO)$^c$ & N(C$^{18}$O)$^c$ & N(C$^{17}$O)$^c$ & N($^{13}$C$^{18}$O)$^c$ & $A_V$$^d$ & N(C$^{18}$O)/N($^{13}$C$^{18}$O) \\
(km\,s$^{-1}$) & (km\,s$^{-1}$) &  10$^{13}$\,cm$^{-2}$ &10$^{13}$\,cm$^{-2}$  &10$^{13}$\,cm$^{-2}$ &10$^{13}$\,cm$^{-2}$ & (mag) & \\
\hline
5.35      &  0.40$^e$& 430        &   50 &  5 & 0.7& 3.5       & 71\\
5.70      &  0.30$^e$& 280        &  150 & 41 & 2.7& 7.9       & 57\\ 
5.90      &  0.30$^e$&  22        &   11 & 2.5& 0.3& 1.9       & 37\\
6.25      &  0.35$^f$& 350        &  100 & 25 & 1.5& 5.5       & 67\\
6.75$^g$  &  0.60    & 130        &      &    &    &1.5$^h$   &\\
\hline
\end{tabular}
\tablefoot{
\tablefoottext{a}{Velocity component.}
\tablefoottext{b}{Adopted line width.}
\tablefoottext{c}{Derived column density in units of 10$^{13}$ cm$^{-2}$. Lines are assumed to be thermalized at the kinetic temperature of the cloud ($T_k$\,=\,9\,K, see Sect.\,\ref{sec:Ancilliary}). The uncertainty of these values is 10\,\%.}
\tablefoottext{d}{Visual absorption derived from the relation obtained in HCL2 
$N$(C$^{18}$O)\,=\,2.5\,$\times$\,($A_V$$-$1.5)\,$\times$\,10$^{14}$ cm$^{-2}$ \citep{Cernicharo1987a}.}
\tablefoottext{e}{For this component the linewidth that best fits the $^{13}$CO emission is 0.6 \kms.}
\tablefoottext{f}{For this component the linewidth that best fits the $^{13}$CO emission is 0.5 \kms.}
\tablefoottext{g}{This velocity component is only detected in $^{13}$CO.}
\tablefoottext{h}{The visual absorption for this velocity component was derived from the relation obtained in HCL2 between the velocity-integrated intensity of $^{13}$CO and $A_V$, $W$($^{13}$CO)\,=\,1.4\,$\times$\,($A_V$$-$0.7) K \kms\, \citep{Cernicharo1987a}.}
}
\end{table*}

In order to interpret the spatial distribution of the species studied with the SANCHO maps, it is of high interest to analyse the line profiles of carbon chains and oxygen/sulfur-bearing species. The goal is to have a view of the physical and chemical properties across the cloud. We selected a series of lines from different species and isotopologues that could bring insights into the physical and chemical structure of TMC-1. A detailed analysis of the abundances of all molecules studied in this work (except CS, and CN that
are analysed in Sect. \ref{depletion_CS} and Sect. \ref{low_opacity_CN}, respectively), will be published elsewhere and we focus here only on their spatial and velocity distribution. 

We first discuss the line profiles of selected molecules toward the $CP$  position. 
Figure\,\ref{fig:ANC_SPEC} shows the observed spectra, mainly at 3\,mm, of the emission of CO, $^{13}$CO, C$^{18}$O, 
C$^{17}$O, 
$^{13}$C$^{18}$O, CH$_3$OH, HCO$^+$, DCO$^+$, H$^{13}$CO$^+$, HC$^{18}$O$^+$, SO, $^{34}$SO, S$^{18}$O, $c$-C$_3$H$_2$, H$_2$CCC, HCCCN, and HCC$^{13}$CN. For polyatomic species several transitions were observed in addition to the lines shown in Fig.\,\ref{fig:ANC_SPEC} and all them show similar line profiles (see, e.g., Fig.\,\ref{fig:CH3OH_lines} for methanol). 
All observed lines are indicated in Table \ref{tab:mapped_lines}.
A comparison of the line profiles of CH$_3$OH, H$_2$CO, $c$-C$_3$H$_2$O, H$_2$CCO, CH$_3$CHO, C$^{34}$S, CCS, and HC$_7$N toward the central source has been previously reported by \citet{Soma2015,Soma2018}, CO, $^{13}$CO, C$^{18}$O, HCO$^+$ and H$^{13}$CO$^+$ have been studied by \citet{Cernicharo1987a}, HCO$^+$ and HCN and their isotopologues have been recently mapped and analysed by \citet{Marcelino2026}, and the spatial distribution and excitation conditions of SO has been studied by \citet{Lique2006}. Large scale maps of the Taurus region in the emission of CO and $^{13}$CO, including HCL2 and TMC-1, have been achieved with the FCRAO telescope \citep{Goldsmith2008,Narayanan2008}. 

\subsection{CO line profiles}\label{sec:CO_line_profiles}

The CO line profile shows emission between 3.5 and 9 \kms. The spatial distribution in Heiles' Cloud 2 of this species \citep{Cernicharo1987a} indicates multiple velocity components and clear saturation effects for $A_V$\,$\ge$\,2 mag. These velocity components have been studied at the scale of the Taurus complex by \citet{Goldsmith2008} and correspond, in addition to the gas of HCL2, to the low visual extinction extended halos of all the nearby Taurus clouds. The $^{13}$CO line profile also indicates clear signs of saturation and its emission covers the velocity range between 4.5 and 7.7 \kms. For $A_V$\,$\ge$\,5 mag, the line was found to be saturated by \citet{Cernicharo1987a}, and hence, it cannot be used to trace the column densities in TMC-1. However, the $J$\,=\,1-0 lines of three rare isotopologues C$^{18}$O, C$^{17}$O (using the $F$\,=\,5/2-5/2 transition which is not blended with the other hyperfine components), and $^{13}$C$^{18}$O correspond mostly to optically thin emission and are well suited to analyze the CO velocity profile. The three rare isotopologues of CO show two main velocity components in TMC-1 at 5.7 and 6.25 \kms\, with some weak emission at 5.35 \kms\, 
and a clear decrease of the emission at 5.8-5.9 \kms\  \citep[the velocity range for carbon chains and cyanopolyynes as derived by][]{Cernicharo2020a}. The emission seen for $^{13}$CO at 6.75 \kms\ is not detected in the other rare isotopologues of CO. The two velocities 5.7 and 6.25 \kms are the blue and red filaments ($BF$ and $RF$, respectively) presented in the maps of Sect.\,\ref{Maps}. They correspond to the low and high velocity components of \citet{Soma2015}.

Assuming that CO and its isotopologues are thermalized at the kinetic temperature of the cloud (9 K; \citealt{Agundez2023b}), we estimate that in order to reproduce their observed line profiles, the 5.7 \kms\, velocity component needs an opacity of 1.6, 0.7, 0.045, and 0.013 for $^{13}$CO, C$^{18}$O, C$^{17}$O ($F$=5/2-5/2), and $^{13}$C$^{18}$O, respectively. The corresponding column densities are given in Table \ref{CO_columndensities}. Hence, although some radiative trapping could occur for C$^{18}$O, and certainly for $^{13}$CO, the other two rare isotopologues of CO provide the total column density of CO at each velocity component (see Table \ref{CO_columndensities}) and could be used to trace the spatial distribution of the gas. 
The component at 5.83 \kms, which corresponds to the cyanopolyyne source, is marginally detected
for all isotopologues and, hence, difficult to fit taking into account the small velocity differences between 
the various velocity components.
Nevertheless, it has been fitted fixing its velocity to 5.90 \kms (the average between the
two strongest components at 5.7 and 6.25 \kms), and its
linewidth to 0.3 \kms for all isotopologues, except for
$^{13}$CO for which it has been fixed to 0.6 \kms. The derived column densities for this velocity
component, which could be identified with the
cyanopolyyne source, are 
typically 5-15 times lower than those of the gas surrounding it. This effect 
will be discussed in Sect. \ref{depletion_CO} as due to a significant depletion of CO into the dust grains.
It is worth noting that the $^{13}$CO line profile is well reproduced when an additional component at 6.75 \kms\ is added (i.e., 5.35, 5.7, 5.90, 6.25, and 6.75 \kms, see Fig.\,\ref{fig:ANC_SPEC}). However, the derived column densities for this isotopologue have to be considered as lower limits due to the opacity problems of the line. From the $N$(C$^{18}$O)/$A_V$ relation obtained in HCL2 by \cite{Cernicharo1987a}, we derived visual absorptions of 3.5, 7.9, 1.9, and 5.5 mag for the 5.35, 5.7, 5.90, and 6.25 \kms\ velocity components, respectively. Our estimation for the column density of C$^{17}$O for the red component is in excellent agreement with the value reported by \citet{Soma2015}. However, for the blue component our estimation is a factor of 1.7 lower than theirs, probably because their reference position is shifted with respect to ours by $\Delta\alpha$=+13$''$ and, hence, our pointing position enters directly in the central zone of CO depletion (see Sect.\,\ref{depletion_CO}).

From the calculated column densities of C$^{18}$O and $^{13}$C$^{18}$O we derive $^{12}$C/$^{13}$C abundance ratios of 
71, 57, 37, and 67 for the velocities of 5.35, 5.7, 5.90, and 6.25 \kms, respectively (see Table \ref{CO_columndensities}). The averaged $^{12}$C/$^{13}$C abundance ratio for all TMC-1 cloud velocities as derived from CO is 60\,$\pm$\,12, which suggests a moderate $^{13}$C fractionation enrichment for CO compared with the isotopic abundances derived from HC$_3$N, HCCNC, and HNCCC, which are between 90 and 100, except for HCC$^{13}$CN which provides a ratio of 69\,$\pm$\,1 \citep{Tercero2024,Cernicharo2024b}.

\subsection{Other species}
A study of the line profiles for other species is provided in Appendix \ref{app:line_profiles} which includes a detailed analysis of the methanol lines (see Appendix \ref{app:methanol}), of CS (see Appendix \ref{app:two_layer}), and
CN (see Appendix \ref{low_opacity_CN}). Additional molecular emission maps 
and line intesity analysis for several species, including S-bearing species, anions, cations, cyanopolyynes and
radicals have been presented by \citet{Tercero2024}, \citet{Fuentetaja2025}, \citet{Marcelino2026}, and
\citet{Cernicharo2026b}.

\begin{figure*}
\centering
\includegraphics[width=0.965\textwidth]{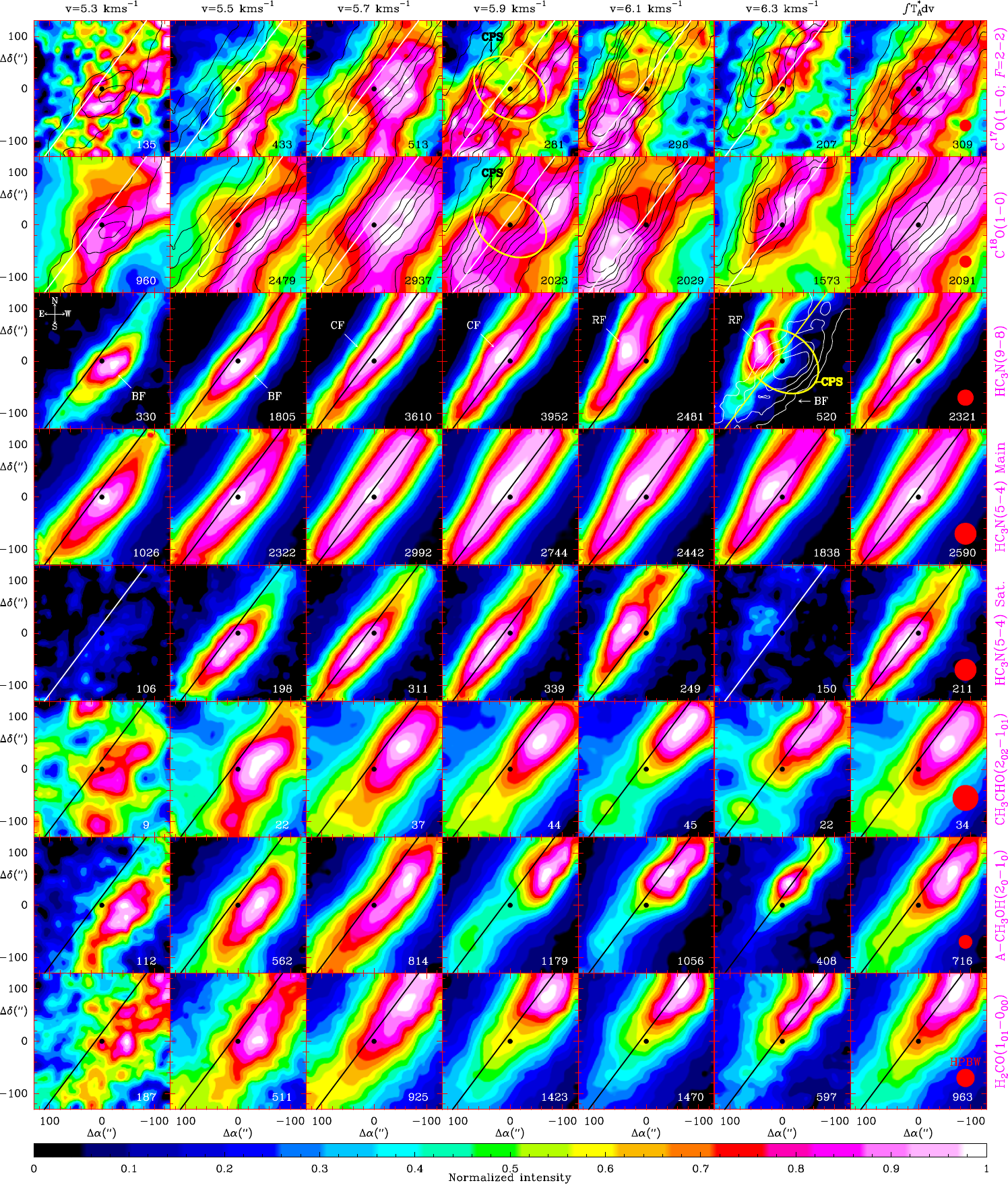} 
\caption{Spatial distribution of the emission ($T_A$$^*$ in mK) as a function of the velocity of different species around the $CP$ of TMC-1 (indicated by a black dot in each panel). The last panel of each row represents the velocity-integrated intensity of the molecule (in mK km\,s$^{-1}$). The first row corresponds to the hyperfine $F$\,=\,2-2 component of C$^{17}$O $J$\,=\,1-0, which is optically thin and, hence, a proxy of the total gas column density. The other panels correspond to the species indicated at the right side of the integrated intensity map. Each map was normalized to its maximum emission indicated in white or black in the bottom-right side of each panel. Over the velocity color maps of C$^{17}$O and C$^{18}$O we show the contours of the intensity of HC$_3$N $J$\,=\,9-8 at the same velocities (first contour 100\,mK with step of 100, 350, 750, 750, 350 and 100 mK from left to right). The integrated intensity contours of HC$_3$N are also superimposed on the C$^{17}$O and C$^{18}$O integrated intensity maps (first contour and step of 250 and 450 mK \kms, respectively). The HPBW of the telescope is shown as a red circle in the integrated intensity panels. The diagonal line (black or white) corresponds to the orientation of the emission of HC$_3$N at 5.7 and 5.9 \kms\, (the $CF$ filament, see Sect.\,\ref{velocity}). The cyanopolyyne source ($CPS$) is delineated as a yellow ellipse in the panel of C$^{17}$O and C$^{18}$O corresponding to $v$\,=\,5.9 \kms\, and over the HC$_3$N $J$\,=\,9-8 color map corresponding to $v$\,=\,6.3 \kms\, (the $RF$ filament; see text); this panel also shows in white contours the emission of HC$_3$N at $v$\,=\,5.3 \kms\, (the $BF$ filament; see text).}
\label{fig:ANC_maps}
\end{figure*}

\begin{figure*}
\centering
\includegraphics[width=0.98\textwidth]{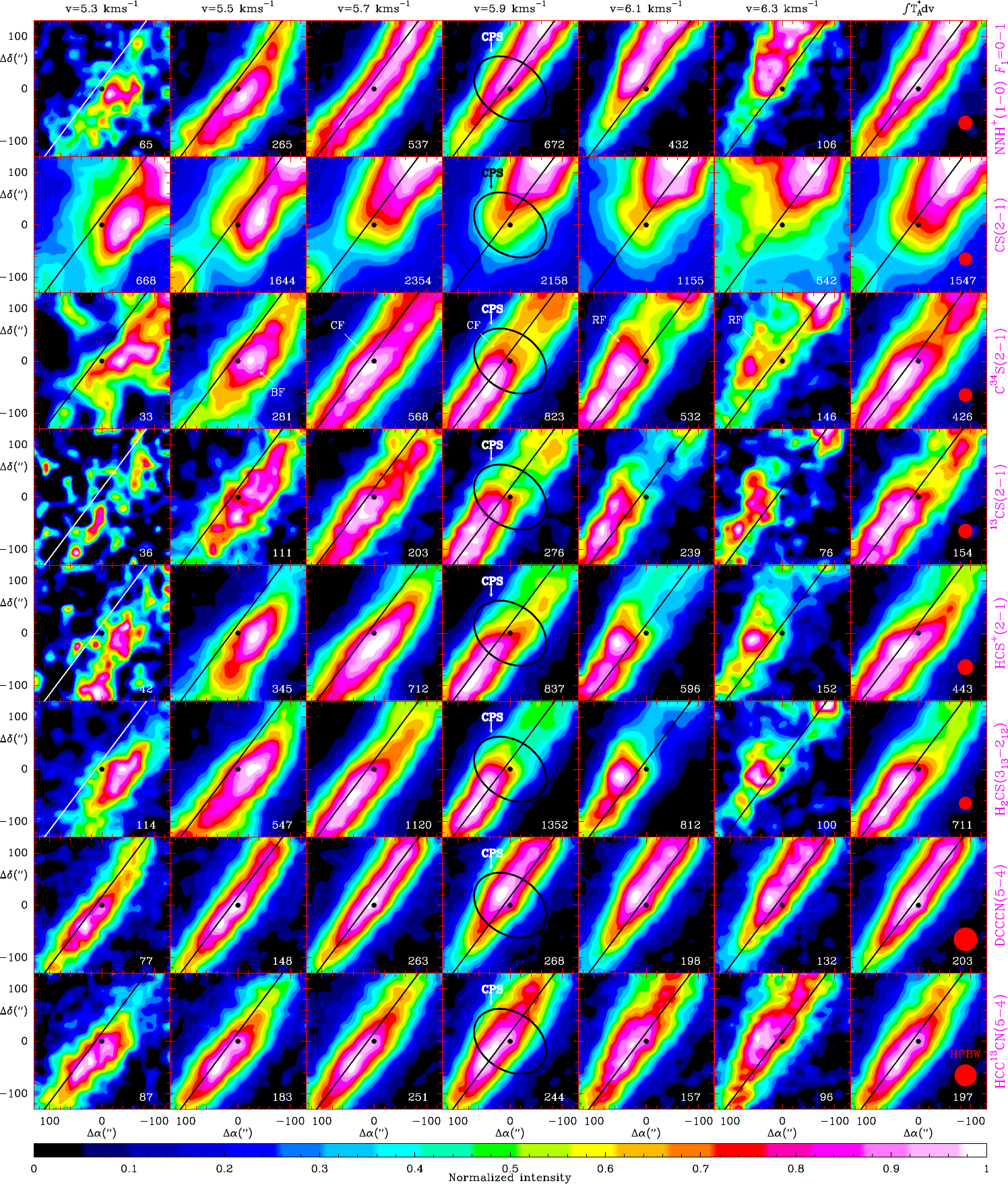}
\caption{Comparison of the spatial distribution at different velocities of the emission of the $J$\,=\,1-0 $F_1$\,=\,0-1 line of NNH$^+$ (this transition has three additional components that are predicted at the same frequency) with that of the $J$\,=\,2-1 line of CS, C$^{34}$S, $^{13}$CS, and HCS$^+$. The plot also shows the spatial distribution of selected lines of H$_2$CS, DCCCN, and HCC$^{13}$CN. Similar to the maps in Fig.\,\ref{fig:ANC_maps}, the emission at each velocity has been normalized to its maximum emission (indicated in white at the right-bottom side of each panel). The black dot corresponds to the $CP$  position. The units for the intensity maximum in each panel correspond to $T_A$$^*$ in mK and are indicated in white. The last panel of each row corresponds to the integrated intensity in mK\,\kms, which also shows the HPBW of the telescope as a red circle at the right-bottom. The cyanopolyyne source, $CPS$ , is indicated in the panels corresponding to $v$\,=\,5.9\,\kms. The black/white line corresponds to the orientation of the emission at 5.7 and 5.9 \kms\, of HC$_3$N, among other species studied in this work (the $CF$ filament, see text and Fig.\,\ref{fig:ANC_maps}).} 
\label{fig:ANC_maps_2}
\end{figure*}

\section{The spatial distribution of O-, C-, and S-bearing species}\label{Maps}
\subsection{Velocity structure}\label{velocity}

In order to have a better view of the gas distribution at each velocity, and to follow in detail possible radiative trapping effects that could affect some molecules, we show in Fig.\,\ref{fig:ANC_maps} the spatial distribution of the emission at velocities between 5.3 and 6.3 \kms\ for the species C$^{17}$O, C$^{18}$O, HC$_3$N $J$\,=\,9-8 and $J$\,=\,5-4 (main hyperfine components which are probably optically thick, and the weak hyperfine satellite lines which are certainly optically thin), CH$_3$CHO, CH$_3$OH and H$_2$CO. In Fig.\,\ref{fig:ANC_maps_2} we show the same for NNH$^+$, CS, C$^{34}$S, $^{13}$CS, H$_2$CS, HCS$^+$, DCCCN, and HCC$^{13}$CN. Superimposed on the C$^{17}$O and C$^{18}$O color maps of Fig.\,\ref{fig:ANC_maps} we show in black contours the emission at the same velocity of the HC$_3$N $J$\,=\,9-8 line, which allows us to compare high- and low-density gas and molecular depletion of species such as CO and CS along the cyanopolyyne filament.

The integrated intensity maps of CS and its isotopologues have been presented by \cite{Fuentetaja2025}. For DC$_3$N $J$\,=\,5-4 and HCC$^{13}$CN $J$\,=\,5-4 integrated intensity maps have been shown by \citet{Tercero2024}, where velocity-integrated intensity maps of the remaining isotopologues of HC$_3$N were also shown. We note that maps of C$_6$H$_5$CN (together with cyanopolyynes, radicals, and other species) have been already shown by \cite{Cernicharo2023}.

Our central position, the classical $CP$  peak (denoted as a black circle in each panel of Figures\,\ref{fig:ANC_maps}, \ref{fig:ANC_maps_2}, and \ref{fig:ANC_maps_3}), was selected from previous maps of TMC-1 \citep{Cernicharo1987a,Pratap1997,Dickens2001,Fosse2001,Lique2006} and coincides with the position used in the line survey performed with the IRAM\,30m telescope \citep{Marcelino2007,Marcelino2009,Cernicharo2012b}, and with that of QUIJOTE. However, the central position used by \citet{Soma2015} is at $\Delta\alpha$=+13$''$ from our CP. In \cite{Langer1995}, the reference position of the maps is at $\Delta\delta$=$-$55$''$ from our CP.

The new sensitive maps shown in this work indicate that the $CP$  position does not always correspond to the intensity maximum, which depends on the molecule, the line, and the velocity component. Similar results were found for the spatial distribution of the velocity-integrated intensity of many molecular lines with data from the SANCHO project \citep{Cernicharo2023,Tercero2024,Fuentetaja2025,Marcelino2026}. We however note that the variation of the intensity between the maximum value at that at the $CP$  position is less than 20\%, even in the high angular resolution maps taken with the IRAM\,30m telescope. This clearly indicates that although some clumpiness is seen, these clumps are partially resolved in our maps producing  a rather flat intensity distribution along the TMC-1 filament.

To follow more clearly the spatial and velocity structure of TMC-1, each panel of Figures\,\ref{fig:ANC_maps} and \ref{fig:ANC_maps_2} shows a black (or white) line that traces the orientation of the emission of HC$_3$N and NNH$^+$ at 5.70 and 5.90 km\,s$^{-1}$. This black/white line traces a filament (hereafter referred to as $CF$, the cyanopolyyne filament). At velocities larger than 5.9 km\,s$^{-1}$ (the red filament, $RF$) and lower than 5.7 \kms\ (the blue filament, $BF$), the emission of these species is found above and below this line, respectively. The $RF$, $CF$ and $BF$ filaments are indicated by white labels in the row corresponding to the emission of HC$_3$N ($J$\,=\,9-8) in Fig.\,\ref{fig:ANC_maps}. The $BF$ and $RF$ filaments trace two arcs South-North(East) and South-North(West), respectively. Both the blue (white contours), and the red (color map) filaments are shown together in the panel of HC$_3$N corresponding to the velocity of 6.3 \kms\, where it is seen that the separation in space between these filaments is less pronounced toward the south. The strongest emission for all carbon-bearing species, the $CF$ filament (black or white line in the panels of Figures\,\ref{fig:ANC_maps} and \ref{fig:ANC_maps_2}), is found between the two arcs. Toward the South, the spatial separation between the $RF$ and $BF$ filaments is less pronounced but still present. Since the HC$_3$N $J$\,=\,9-8 line is certainly optically thick, we also show in this figure the emission of the main component of the $J$=5-4 line (which represents 97.3\,\% of the emission of this line), and that of its weak hyperfine satellite lines ($F$\,=\,5-5 and $F$\,=\,4-4) which, when added, represent 2.7\,\% of the emission of the main component. These satellite lines are optically thin and trace the high-density gas in the cloud and the highest column densities in HC$_3$N. The corresponding hyperfine lines of the $J$\,=\,9-8 transition are too weak and were not detected with the SANCHO maps taken with the IRAM\,30m telescope. Indeed, the gas traced by the weak hyperfine components of the HC$_3$N $J$\,=\,5-4 line appears to be surrounded by an extended halo traced by the main component of this line. 

\begin{figure}
\centering
\includegraphics[width=0.95\columnwidth]{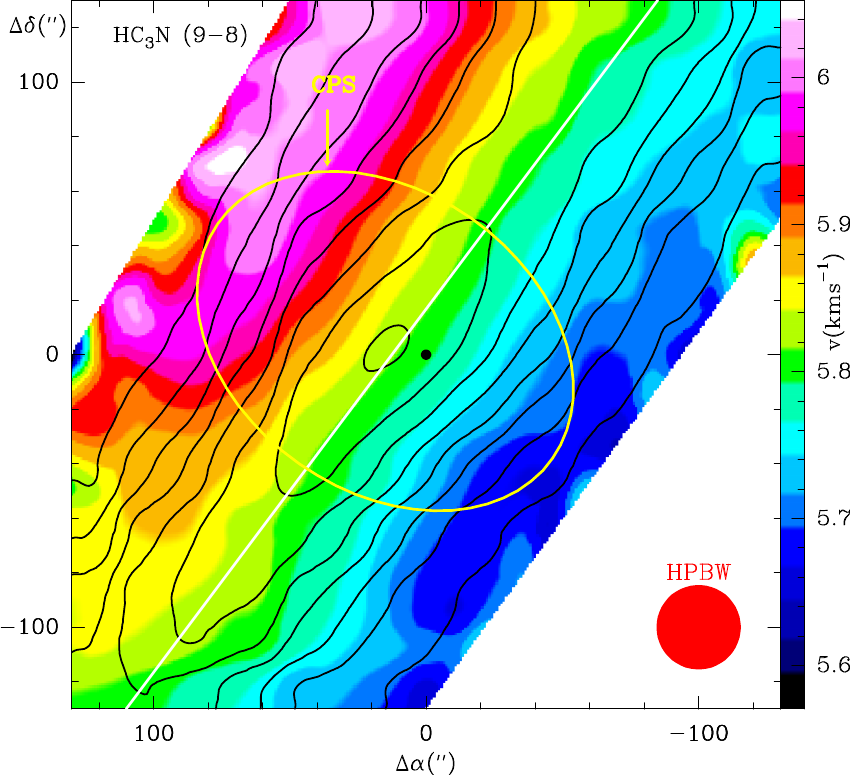}
\caption{Spatial distribution of the velocity centroid of HC$_3$N $J$\,=\,9-8 (in colors) and velocity-integrated intensity of the same line in black contours (0.2 to 2.5 by 0.3 K\,\kms). The white straight line corresponds to the orientation of the $CF$ filament discussed in the text. The HPBW of the telescope is shown as a red circle. The black dot indicates the $CP$  position. The yellow ellipse centered at $\Delta\alpha$=15$''$ and $\Delta\delta$=5$''$ with respect to the $CP$  shows the cyanopolyyne source ($CPS$, see text), whose major and minor axis are 150$''$ and 110$''$, respectively.} 
\label{fig:HC3N_moments}
\end{figure}

The $RF$, $CF$, and $BF$ filaments are clearly seen in most of the molecules presented in Figures\,\ref{fig:ANC_maps} and \ref{fig:ANC_maps_2}. Except for CO, CS, CH$_3$CHO, H$_2$CO and CH$_3$OH, which will be discussed in the next sections, all C-bearing species studied in this work show the same spatial 
behavior for the different cloud velocity components (see Figures\,\ref{fig:ANC_maps} to \ref{fig:ANC_maps_3}).

Along the $BF$, $RF$, and $CF$ filaments several clumps can be identified at different velocities. Higher angular resolution observations could even indicate the presence of additional structures in these clumps. The emission peak of the C-bearing species is close to the $CP$ but oscillates South-West 
(blue velocities) and North-East (red velocities) of the $CP$  position following a line perpendicular to the orientation of the $CF$ filament. This emission is enclosed in an ellipse denoted hereafter as $CPS$ (cyanopolyyne source). It is shown in yellow in the panel at $v$\,=\,5.9 \kms\ of C$^{17}$O and C$^{18}$O in Fig.\,\ref{fig:ANC_maps} and it contains all the strong emission at all velocities (above 20\,\% of the maximum intensity) of the species studied in this work. It corresponds to an ellipse of major and minor axis of 150$''$ and 110$''$ respectively. It is also shown in the panel with $v$\,=\,6.3\,\kms of HC$_3$N $J$\,=\,9-8 of the same figure, and in all panels at $v$\,=\,5.9\,\kms\ of Fig.\,\ref{fig:ANC_maps_2}. The position angle of the $CPS$ source is 35$^{\circ}$, i.e., nearly perpendicular to the $CF$ filament (black or white lines in the panels of Figures\,\ref{fig:ANC_maps} and \ref{fig:ANC_maps_2}). In the direction perpendicular to the filament, the size of the integrated emission at half power intensity varies between 70$''$ and 80$''$ depending on the position selected for the cut and on the observed transition. Hence, taking into account these results, the adopted source size in our previous works with QUIJOTE data seems to be a reasonably good approximation. Nevertheless, three elliptic sources at velocities 5.3-5.7 \kms\ ($BF$), 5.7-5.9 \kms\ ($CF$), and 5.9-6.3 \kms\ ($RF$) with major and minor axis of 100$''$ and 70-80$''$, respectively, could be a better choice to reproduce the observed velocity distribution. We verified that the dilution factors adopting a circular source or an elliptic one do not introduce significant errors in the determination of molecular column densities.

If we restrict our view to the $CPS$, it could be tempting to assign the observed velocity field to a rotating disk. However, the main emission inside the $CPS$ belong to the $CF$ filament with the $RF$ and $BF$ filaments being adjacent to it (right and left shifted) and having a lower intensity. Hence, the most plausible explanation is that the maps of carbon-bearing species shown in Fig.\,\ref{fig:ANC_maps} correspond to a filament (the $CF$ filament) with a velocity gradient perpendicular to it. To strength this hypothesis we show in Fig.\,\ref{fig:HC3N_moments} the centroid velocity of HC$_3$N $J$\,=\,9-8 (color map), together with its integrated intensity (black contours). This velocity gradient is present over the whole structure of the TMC-1 cloud. In this spatial view of TMC-1 the $BF$ and $RF$ filaments correspond to the edges of the $CF$ filament and have a velocity difference of $\sim$1 \kms. As both filaments are separated by $\sim$100$''$, the velocity gradient perpendicular to the  filament is 6.7\,$\times$\,10$^{-18}$ km\,s$^{-1}$\,cm$^{-1}$. Large scale mapping of HCL2 (one square degree) in several molecules by \citet{Cernicharo1987a} indicate that the cloud is dominated by two large filamentary structures with a velocity distribution peaking at 5.3 and 6.1 \kms. Hence, the TMC-1 fragment of HCL2 inherits, in some way, the large scale velocity structure of the cloud. The spatial distribution of the different velocities in TMC-1 have been also discussed in detail by \citet{Lique2006}.

\subsection{Depletion of carbon monoxide}\label{depletion_CO}

The emission of C$^{17}$O and C$^{18}$O at several velocities is shown in the first two rows of Fig.\,\ref{fig:ANC_maps}. This emission is clearly deficient along the $CF$ filament at several velocities as compared with other species, in particular with HC$_3$N which is a proxy of the emission of cyanopolyynes. The intensity of the emission of HC$_3$N at different velocities is superimposed on the C$^{17}$O and C$^{18}$O color maps as black contours. The spatial distribution of the velocity-integrated intensity of both isotopologues is concentrated South-West of the $CF$ filament with a considerably decrease of the intensity over the $CF$, particularly toward the East above the black/white line denoting this filament, and where HC$_3$N and other carbon chains are concentrated (see Fig.\,\ref{fig:ANC_maps} and \ref{fig:ANC_maps_2}). Considering the velocity maps, this spatial shift appears at velocities 5.5 and 5.7 \kms\ with emission peaks shifted by more than 70$''$ (more than three HPBW). At other velocities, there are some regions of the map that show little emission of C$^{17}$O and C$^{18}$O. This is the case of the South-East region of the $CF$ filament at $v$\,=\,5.7 \kms. Moreover, the central $CPS$ region at 5.9 \kms\ is practically empty of emission in C$^{17}$O and shows a significant decrease of C$^{18}$O emission. At the velocity 6.1\,\kms\, the emission of the isotopologues of CO is shifted toward the South of the $CF$ filament. However, the spatial shift between the $BF$ and $RF$ filaments observed in HC$_3$N do not have a clear counterpart in these two isotopologues of CO. In fact, the $RF$ filament is not traced at all by any of the isotopologues of CO. This suggests a significant depletion of carbon monoxide in the densest part of the $BF$, $CF$, and $RF$ filaments. We note that the velocity of the $CF$ filament is 5.83\,\kms \citep{Cernicharo2020a} while the isotopologues of CO, including the extremely rare $^{13}$C$^{18}$O, show a significant lack of emission at this velocity (see Sect.\,\ref{sec:CO_line_profiles}). Hence, most of the observed emission of C$^{17}$O and C$^{18}$O corresponds probably to the low density envelope of the $CF$ filament. This behaviour of CO being the skin of the cloud is clearly shown by the comparison of the $BF$ as defined by HC$_3$N $J$\,=\,9-8 contours superimposed on the $v$\,=\,5.5 \kms color map of both isotopologues of CO in Fig.\,\ref{fig:ANC_maps}. Previous maps of C$^{18}$O $J$\,=\,1-0 by \citet{Langer1995} have a sampling of 1 arcmin and were taken with a beam of 100$''$ compared with our beam of 23$''$, and could not provide a detailed view of the spatial distribution of the emission from the CO isotopologues.

Previous chemical models of TMC-1 strongly support a C/O abundance ratio $\ge$1 to explain the observed abundances of the carbon chains \citep[][see also \citealt{Turner2000}]{Cernicharo2021h}. If CO is depleted, then it is also reasonable to assume that H$_2$O will be also mainly on the grains. Hence, the two main molecular carriers of oxygen should be condensed on the grains, i.e., oxygen will depleted in the gas-phase along the $CF$ filament. It is important to take into account that CO freeze-out could lead to the formation of complex organic molecules (COMs) on dust grains over timescales of a few 10$^4$ 
years \citep{Tielens1982,Watanabe2002,Cuppen2009,Fuchs2009}. These molecules, when totally released to the gas phase due to warm-up of the grains during star formation, will dominate the millimeter and submillimeter spectra of hot cores and hot corinos.

The determination of the depletion factor of CO is rather difficult due to the spatial shift of the emission from these isotopologues with respect to that of the velocity of the $CF$ filament
(see Sect. \ref{sec:CO_line_profiles}). We estimate that the difference in the emission of the isotopologues of CO over the $CF$ filament at the velocity of the dense gas, and their emission maxima, always South-West of the $CF$ filament, suggest a moderate CO depletion of $\sim$10 in the densest parts of TMC-1. This value is estimated from the contrast between the two velocity peaks of the C$^{17}$O emission and the emission at 5.83 \kms\, in Fig.\,\ref{fig:ANC_maps}. We also considered the intensity ratio of C$^{17}$O between the edges and the center of the $CPS$ source. 
The analysis of the line profiles of CO and its isotopologues given in Sect. \ref{sec:CO_line_profiles} also suggests a depletion of $\sim$5-15.
 
The depletion factor of CO has been measured in several clouds showing a large range of values, ranging from 100 in B68 \citep{Bergin2002}, 2-10 in L694 \citep[][and references therein]{Taillard2025}, and $\sim$10 in infrared dark clouds \citep{Cosentino2026}.
The spatial distribution of CO with respect to other species in TMC-1 follows in many aspects that of other cold clouds. What makes special TMC-1 is that its density is 1-2$\times$10$^{4}$ cm$^{-3}$ and we do not expect significant freeze-out of CO at these moderate densities \citep[see, e.g., the models of][]{Taillard2025}. Nevertheless, the depletion of CO has important consequences on the chemistry of hydrocarbons and certainly in that of PAHs \citep{Cernicharo2021a} as the C/O abundace ratio in the gas phase has a strong change when depletion removes a significant fraction of CO. Depletions of $\sim$2-3 are expected for the densities of TMC-1 (see, e.g., Fig.\,9 of \citealt{Taillard2025}, and Fig.\,6 of \citealt{Cosentino2026}). A strong exponential dependence on the dust temperature has been found for the depletion of CO by \citet{Cosentino2026}. Taking into account that the kinetic temperature in TMC-1 is $\sim$9\,K \citep{Agundez2023b}, we could expect an enhancement of the CO depletion even at the moderate densities of this source. Moreover, dust properties such as grain size could also take a role in enhancing CO depletion. Studies of the dust emission in Taurus filaments suggest an important change in dust properties between the external and inner regions of these filaments \citep{Ysard2013}. The study of the dust opacity at 250 $\mu$m in several dark clouds in  Ophiuchus by \citet{Li2024} also suggests grain growth when density increases. Hence, CO depletion has to be considered as a dynamical process during the collapse history of dense cores such as TMC-1.

\subsection{S-bearing species and depletion of carbon monosulfide} \label{depletion_CS}

The spatial distribution of the velocity integrated intensity of several S-bearing species has been presented by \citet{Fuentetaja2025}. In Fig.\,\ref{fig:ANC_maps_2} we show the spatial distribution of CS, C$^{34}$S, $^{13}$CS, HCS$^+$, H$_2$CS at different velocities. NNH$^+$, DCCCN and HCC$^{13}$CN are also shown in this figure for comparison purposes.

As discussed by \citet{Fuentetaja2025}, the CS abundance would be highly underestimated using the $J$\,=\,1–0 and $J$\,=\,2–1 lines, which are optically thick and, hence, cannot be used to derive physical parameters and isotopic abundances. These lines are so optically thick that they trace only the external layers of the cloud (see below). The line opacity problems are more clearly seen in the comparison of the velocity maps of the intensity of CS and its $^{13}$CS and C$^{34}$S isotopologues (see Fig.\,\ref{fig:ANC_maps_2}). The intensity of CS peaks toward the Norht-West of the maps and is shifted along the main axis of TMC-1 by more than 2 arc minutes with respect the emission of its $^{13}$CS and C$^{34}$S isotopologues, which however, coincide with the gas around the $CP$  and the $CPS$. Moreover, the emission of CS seems to decline toward the zone in which the isotopologues are most intense which indicates strong radiative trapping effects on the lines of the main isotopologue. Toward the CS peak at the North-West the weakness of the emission from the isotopologues suggests a strong decreases of their column density, i.e., a strong depletion of CS. 

The radiative trapping effects discussed above affect both the $J$\,=\,2-1 line (show here) and the $J$\,=\,1-0 line (shown by \citealt{Fuentetaja2025}). HCS$^+$ and H$_2$CS follow the spatial distribution of the CS isotopologues at all velocities, showing a clear deficiency in the emission toward the North-West part of the $CF$. The contrast in emission in the isotopologues between the South and North parts of the $CF$ is $\sim$5. Hence, we have two different effects to be taken into account when analyzing CS. First, the line opacity of the main isotopologue prevents the use of the emission distribution to derive the abundance distribution. The second problem is related to depletion, as traced by the rare isotopologues. Both C$^{34}$S and $^{13}$CS show that most of the emission is arising from the South of the $CF$ with a significant decrease of their emission toward the North-West, where the emission of the main isotopologue peaks. The simplest explanation is that CS is not depleted in the South and that the isotopologues trace the dense gas. Moreover, the inner and dense layer responsible for the emission of the isotopologues has not a significant column density in front of it to affect their emerging intensities. However, for CS the opacity is huge in these external regions and the CS photons coming from the densest regions (where 
the observed $^{13}$CS and C$^{34}$S lines are formed), cannot escape from the cloud. Toward the North-West the CS emission traces the gas of the $CF$ but its abundance is much lower, as traced by the isotopologues which have little emission in these zones. Hence, although contrary to what the visual inspection of the CS maps seems to indicate, CS is depleted toward the Nort-West where its emission is peaking but where the emission of the isotopologues is weak compared with the South-West part of the $CF$ filament. As a support to that view, the maps of HCS$^+$ (chemically linked to CS) and H$_2$CS have a similar behaviour to that of the isotopologues of CS. However, the emission of NNH$^+$ $J$\,=\,1-0 $F_1$\,=\,0-1 at 5.9 \kms\, peaks at the North-West of the $CF$ filament, i.e., in the zone where CO and CS are depleted. NNH$^+$ follows nicely the spatial distribution at all velocities of DCCCN (see Fig.\,\ref{fig:ANC_maps_2}), whose emission appears mainly toward the North of the $CF$ filament. 
A detailed analysis of the NNH$^+$ emission has been provided by \citet{Cernicharo2026b}.
A two layer model for CS that accounts for these opacity problems is explained in
Appendix \ref{app:two_layer}.
Similar opacity effects have been also observed for CN. This species is analysed in detail in Appendix \ref{low_opacity_CN}.

\begin{figure*}
\centering
\includegraphics[width=0.98\textwidth]{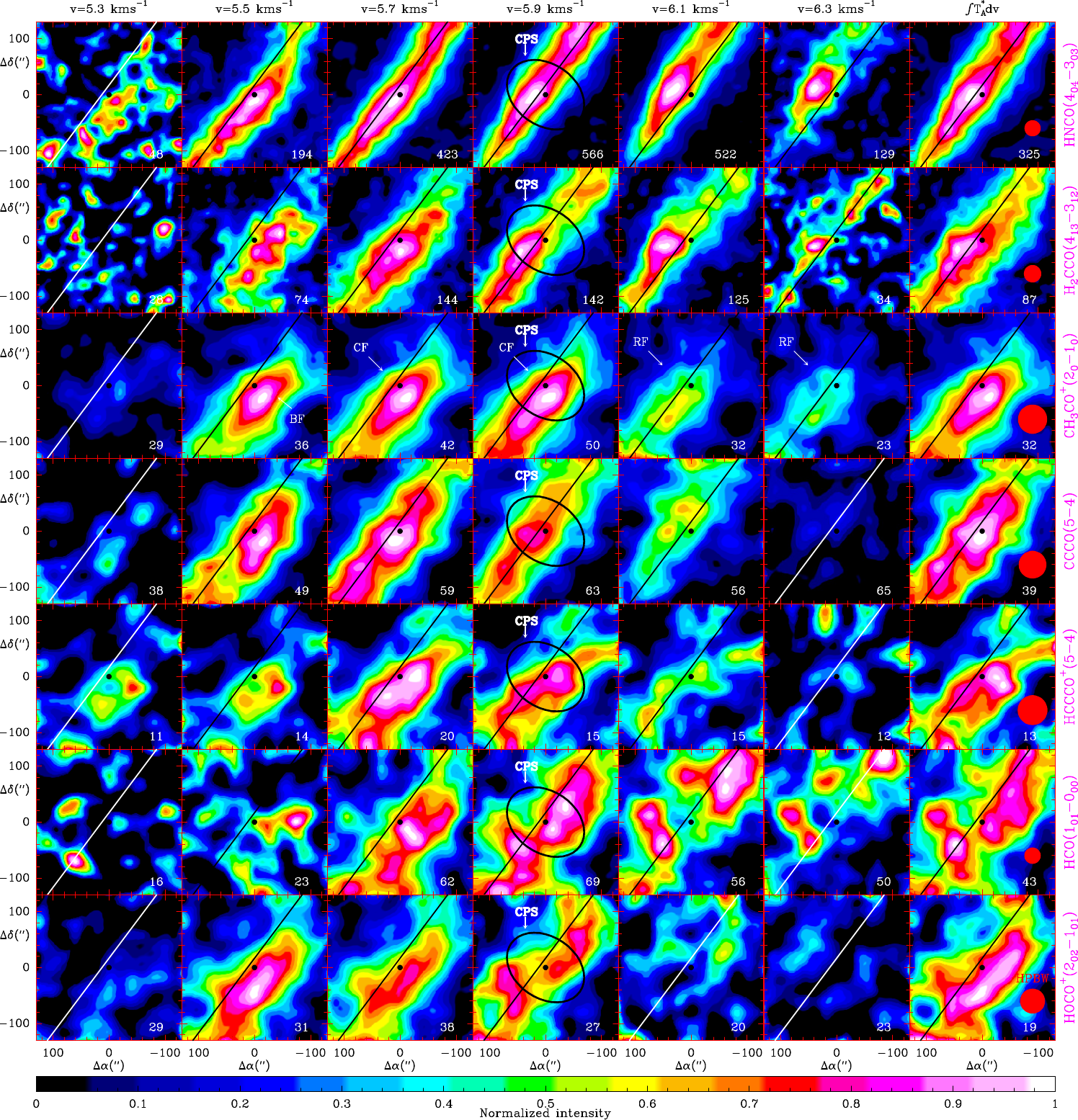}
\caption{Comparison of the spatial distribution at different velocities of the emission of HNCO, H$_2$CCO, CH$_3$CO$^+$, CCCO, HCCCO$^+$, HCO and HOCO$^+$. Similar to the maps in Fig.\,\ref{fig:ANC_maps}, the emission ($T_A^*$ in mK) at each velocity has been normalized to its maximum emission which is indicated in white at the right-bottom side of each panel. The last panel of each row corresponds to the integrated intensity in mK\,\kms. The black dot corresponds to the $CP$  position. The HPBW of the telescope is shown as a red circle in the right-bottom panel. The cyanopolyyne source, $CPS$ , is indicated in the panels corresponding to $v$\,=\,5.9\,\kms. The black/white line corresponds to the orientation of the emission at 5.7 and 5.9 \kms\, of HC$_3$N, among other species studied in this work (the $CF$ filament, see text and Fig.\,\ref{fig:ANC_maps}).} 
\label{fig:ANC_maps_3}
\end{figure*}

\subsection{The spatial distribution of O-bearing species}\label{O-bearing}

In Fig.\,\ref{fig:ANC_maps} we show the spatial distribution at different velocities of the O-bearing species CH$_3$CHO, CH$_3$OH, and H$_2$CO. These three species have a spatial distribution very different from that of the species discussed until now. The observed spatial emission follows more closely that of $^{17}$CO and C$^{18}$O at all velocities. For example, at $v$\,=\,5.7 and 5.9\,\kms\ the emission of CH$_3$CHO, CH$_3$OH and H$_2$CO peaks at a position $\sim$100$''$ North-West from that of HC$_3$N, as already pointed out by \citet{Soma2015} for methanol. For the red filament ($RF$), the emission of these O-bearing molecules is at the right of the $CF$ filament, while for cyanopolyynes and radicals the emission is at the left of $CF$. This North-West emission source in CH$_3$OH was mapped by \citet{Soma2015} and compared to that of C$^{34}$S. In our maps not only methanol, but also CH$_3$CHO and H$_2$CO, appear completely anticorrelated with HC$_3$N and all other C-bearing species shown in Fig.\,\ref{fig:ANC_maps_2}, including that of the isotopologues of CS. In fact these O-bearing species have their emission maxima in the direction of maximum depletion of CS as traced by the isotopologues of this species (see Sect.\,\ref{depletion_CS}). For the velocities corresponding to the blue filament ($BF$), the emission of these three O-bearing species appears to surround that of cyanopolyynes, NNH$^+$, and CN. Moreover, the emission is very weak toward the South-East part of  the $CF$ filament, where CO is also depleted (see Sect.\,\ref{depletion_CO}). Hence, these species that are thought to be mostly formed in the ices of dust grains, appear mainly in the regions where CO is not yet depleted, i.e., in the envelope of the $CF$ filament, as traced by its blue and red components. 

It could be tempting to invoke opacity problems to interpret the spatial behaviour of methanol. Although at velocities 5.9 and 6.1\,\kms the maps look very similar to those of CS and the optically thick hyperfine lines of CN, the fact that we have clearly detected $^{13}$CH$_3$OH (see Fig.\,\ref{fig:CH3OH_lines} and Appendix \,\ref{app:methanol}) with the same line profile than CH$_3$OH indicate that the methanol lines are not very optically thick. Hence, the spatial distribution of the line intensity traces the abundance of methanol rather than radiative trapping problems. \citet{Soma2015} conclude that the different spatial distribution of carbon-chains and methanol is related to the difference in their production process rather than to depletion. However, taking into account the strong CO and CS depletion we found, and the fact that not only methanol, but also CH$_3$CHO and H$_2$CO, are showing the same behavior, we need to consider that in TMC-1 molecular depletion is playing a non-negligible role in the chemistry.

Thanks to the sensitivity of the SANCHO maps we can also study the spatial distribution of other O-bearing species such as HNCO, HOCO$^+$, HCO, HCCCHO, CCCO, HCCCO$^+$, and CH$_3$CO$^+$. These additional maps are shown in Fig.\,\ref{fig:ANC_maps_3}. HNCO follows a spatial distribution at all velocities identical to HC$_3$N and to the optically thin hyperfine lines of CN. Its formation mechanism in the gas phase is considered to be closely related to that of carbon chains, probably involving the dissociative recombination of H$_2$NCO$^+$ \citep{Marcelino2009,Marcelino2010}. Formation on ices implies that NCO formed in the gas is quickly accreted on the grains and that subsequent hydrogenation reactions produce HNCO and its isomers \citep{Quan2010}. Once the grains are heated and molecules start to evaporate, HNCO would be released to the gas phase. In TMC-1 desorption from grains is a slow process
only partially enhanced by non-thermal mechanisms such as chemical desorption, cosmic ray sputtering or photodesorption by secondary UV photons. Hence, if HNCO would have an ice origin, its spatial distribution would be similar to that of CH$_3$OH (or CH$_3$CO and H$_2$CO). This is clearly not the case, as HNCO follows very closely the spatial distribution of carbon chains, CN, and NNH$^+$. Hence, our map shows that gas-phase reactions are the main production mechanism of HNCO in cold starless cores. This is in good agreement with the gas-phase models for cold clouds described by \cite{Marcelino2010}.

Among the species containing O, in addition to HNCO, we show in Fig.\,10 that it is interesting to compare ketene (H$_2$CCO) and the acetyl cation (CH$_3$CO$^+$). None of the two species follow the spatial distribution of H$_2$CO, CH$_3$CO, or CH$_3$OH. They show some similarities in their spatial distribution as it could be expected if the acetyl cation is formed by protonation of ketene \citep{Cernicharo2021g}, but there are significant differences. The three filaments, $BF$, $CF$, and $RF$, are detected in ketene with a similar spatial extent to that of carbon-chains. However, the acetyl cation follows ketene only in the $BF$. All the emission of this species is always below the $CF$ filament (at the right side of the straight line delineating the $CF$ filament). No significant emission of CH$_3$CO$^+$ is found over the $CF$ and the $RF$ filaments. The maps of the velocity integrated intensity of both species are folded with respect to the $CF$ straight line, ketene to the left side, the acetyl cation to the right. Opacity problems are unlikely to occur for both species. Hence, the observed effect is real and indicates a variation of the efficiency of protonation in $CF$ and $RF$ with respect to the blue velocity component. CH$_3$CO$^+$ appears to arise from the single cloud corresponding to the $BF$.

Two additional closely related species, CCCO and HCCCO$^+$, were mapped with SANCHO and are shown in Fig.\,\ref{fig:ANC_maps_3}. Protonated CCCO was detected in TMC-1 with the QUIJOTE line survey by \cite{Cernicharo2020b}. While CCCO has a distribution similar to HNCO and carbon-chains, HCCCO$^+$ has a close spatial distribution to that of CH$_3$CO$^+$. It is curious that both cations have this discrepancy with respect to their neutral precursors. However, such a difference could exit if the abundance of the protonation precursors, i.e., H$_3$$^+$ and HCO$^+$, change in function of the velocity component. A change in the H$_2$ density between the different cloud components would affect the abundance of H$_3$$^+$ and also that of HCO$^+$.

Finally, HCO and protonated CO$_2$ (HOCO$^+$) were also mapped. We found that HCO, an ice-grain chemical product, follows very well the spatial distribution of methanol. HOCO$^+$ appears mainly at the right of the blue filament with some emission over the $CF$. The velocity integrated intensity spatial distribution of both species trace the skin the of $BF$ filament.

\section{Chemical modeling}
\label{sec:Discussion}

\begin{figure*}
\centering
\includegraphics[width=0.95\linewidth]{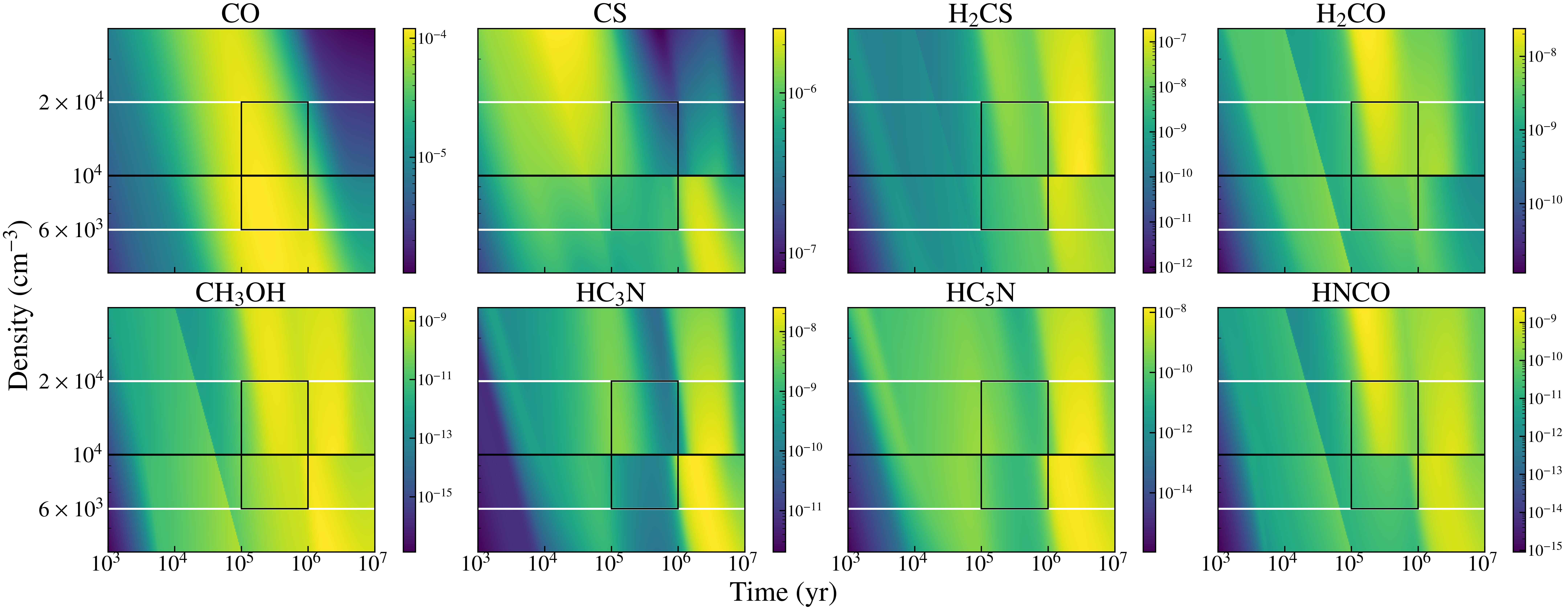}
\caption{Results of the chemical model for selected molecules discussed throughout the manuscript. We include representative depleted species (CO, CS), their grain related molecules (CH$_3$OH, H$_2$CO, H$_2$CS), and species that do not show clear depletion signatures (HC$_3$N and HNCO). HC$_5$N is also included to illustrate the common behavior observed across cyanopolyynes. The white horizontal lines indicate the densities adopted for the envelope (lower line) and core (upper line) models. The black horizontal line marks the density at which the step-function change in visual extinction is applied, while the black rectangle highlights the time window used for the detailed analysis discussed in the text. 
The color bar represents the fractional abundance of the species respect to the total number of H nuclei.}
\label{fig:models_1}
\end{figure*}

\begin{figure*}
\centering
\includegraphics[width=0.95\linewidth]{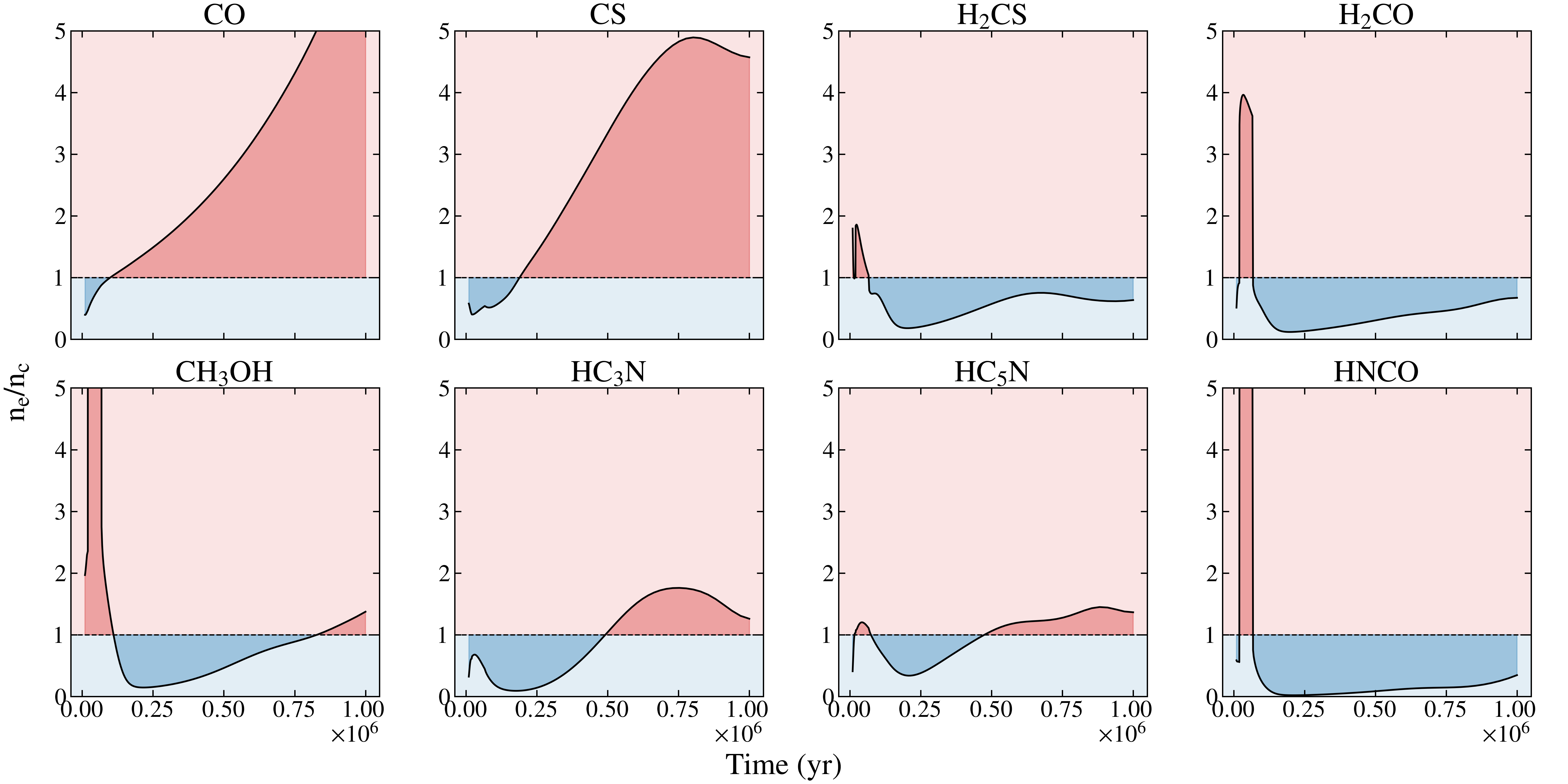}
\caption{Ratio of abundances in the envelope ($n_{\rm model,env}$) and core $n_{\rm model,core}$ for the molecules shown in Fig.\,\ref{fig:models_1} in the 10$^5$-10$^6$ yr time window. A red area under the curve at a given time indicates higher fractional abundance in the envelope whereas a blue area above the curve indicates the opposite.}
\label{fig:models_2}
\end{figure*}

We carried out chemical kinetics models to identify how state-of-the-art reaction networks perform in the prediction of the selective depletion process that we reported here. We used the \texttt{nautilus} three phase gas-grain astrochemical model \citep{Ruaud2016}, coupled to the \texttt{kida.2022}\footnote{https://kida.astrochem-tools.org} gas and grain reaction networks. We chose this model and network to better match the initial conditions of the model reported in \citet{Wakelam2021}. Briefly, Nautilus is a three-phase model in which gas, surface icy mantle, bulk icy mantle, and their interactions are considered. Nautilus solves the kinetic equations for both the gas phase species and the surface species of interstellar dust grains and computes the evolution with time of chemical abundances for a given physical structure. 

For this set of models, to a first approximation we adopted the envelope/core model described in Appendix \ref{app:two_layer}, assumed a constant $T_{\rm g}$\,=\,$T_{\rm d}$\,=\,10 K, and run 2000 models in which the particle density was varied in the range between 4\,$\times$\,10$^{3}$ and 4\,$\times$\,10$^{4}$ cm$^{-3}$. The densities in the core and envelope are estimated to be 2\,$\times$\,10$^{4}$ and 6\,$\times$\,10$^{3}$ cm$^{-3}$, respectively. For simplicity, we adopted a stepwise function for the visual extinction in the simulation of the two-level envelope/core model of the type
\[
\begin{aligned}
A_V = 4.0\,{\rm mag} & \quad \text{if } n < 10^{4}\,{\rm cm}^{-3} \\
A_V = 10.0\,{\rm mag} & \quad \text{if } n > 10^{4}\,{\rm cm}^{-3}.
\end{aligned}
\]
\noindent
where the limit between 4 and 10 mag representd the transition between a somewhat translucent cloud and a completely dark core (see the visual extintion maps of \citet{Cernicharo1987a} and \citet{Navarro2020}).

For the cosmic ray ionization rate, $\zeta$, we adopted the formalism of \cite{Wakelam2021}, which is based on \cite{Neufeld2017} and yields $\sim$5.0\,$\times$\,10$^{-17}$ s$^{-1}$  for $A_V$\,=\,10 mag and $\sim$9.5\,$\times$\,10$^{-17}$ s$^{-1}$ for $A_V$\,=\,4 mag. The elemental abundances are the same reported in Table 1 of \citet{Wakelam2021} for both the envelope and core models. With this simplified model, analogous to the one shown in the observations, we intend to provide a first order approximation to the observational data, postponing a more detailed modeling effort to subsequent works.

\begin{table}
\caption{Observed and modeled column densities for molecules in Figure \ref{fig:models_1}. Two characteristic times are shown, 2$\times$10$^{5}$/ 5$\times$10$^{5}$ years}
\label{tab:columns}
\centering
\small
\begin{tabular}{lccc}
\hline\hline
Molecule & $N_{\rm obs}$ & $N_{\rm model,env}$ & $N_{\rm model,core}$ \\
\hline
CO & 1.7(18) & 2.4(18)/2.2(18) & 1.8(18)/8.4(17) \\
H$_2$CO & 5.0(14) & 4.0(13)/6.4(13) & 3.4(14)/2.2(14) \\
CH$_3$OH & 4.8(13) & 3.2(12)/1.0(13) & 2.2(13)/2.2(13) \\
CS & 1.1(14) & 1.4(16)/1.5(16) & 1.3(16)/4.6(15) \\
H$_2$CS & 3.7(13) & 8.0(13)/1.6(14) & 4.4(14)/2.8(14) \\
HC$_{3}$N & 2.3(14) & 3.6(12)/2.2(12) & 3.8(13)/2.0(12) \\
HC$_{5}$N & 1.8(14) & 1.2(12)/5.8(11) & 3.6(12)/5.4(11) \\
HNCO & 1.1(13) & 5.4(11)/6.4(11) & 2.6(13)/7.4(12) \\
\hline
\end{tabular}
\tablefoot{Column densities are in units of cm$^{-2}$, with the notation A(B) referring to A\,$\times$\,10$^{B}$ taking $N_{\rm obs, \ce{H2}}$=1\,$\times$\,10$^{22}$ cm$^{-2}$. Observational column densities are taken from \cite{Agundez2026}.}
\end{table}

The results for the grid of chemical models are shown in Fig.\,\ref{fig:models_1}. At times between 10$^5$ and 10$^6$ yr, models reflect a mature chemistry where most of the molecules displayed are already present in significant amounts. Assuming a total column density of 10$^{22}$ cm$^{-2}$ for TMC-1 \citep{Cernicharo1987a}, 
and taking the molecular column densities at a characteristic time of 2$\times$10$^{5}$ and 5$\times$10$^{5}$ years \citep[which are the typical times considered for the
chemical evolution of TMC-1, see, e.g.,][]{Agundez2026}, 
we obtain the column densities of each molecule shown in Figure \ref{fig:models_1} and compare it with the latest available column densities for them \citep{Agundez2026}. This is done in Table \ref{tab:columns}. The overall agreement between the predicted and observed abundances ranges from very good for most molecules to poor for cyanopolyynes. However, in this work we are more interested in a qualitative prediction of the column densities. Nonetheless, it is important to mention that a modeling using a more detailed physical structure of the cloud is likely to capture better the prediction of cyanopolyyne abundances. It is meritorious to also mention that \ce{CH3OH}, \ce{H2CS} and \ce{H2CO} are consistently more abundant in the envelope than \ce{HC3N} \ce{HC5N} and HNCO, aligned with the specific presence in the core of the latter molecules, consistent with observations.

In addition to the static picture provided given by the abundances at given times, it is also interesting to investigate how the relative abundances vary over time for $n_{\rm model,env}$ / $n_{\rm model,core}$, as this is the quotient that we can directly compare with observations. Such a quotient is shown in Fig.\,\ref{fig:models_2} in the 10$^5$-10$^6$ yr timeframe for the same molecules already considered throughout this section. Many points of comparison can be established with the observations from this figure. First, our results show a clear depletion of CO and CS in the core, which is more pronounced at t=5$\times$10$^{5}$ years as it is expected from the increased particle density, and depletion timescales. For CO and CS we can attribute the change in the abundance exclusively to depletion onto grains, allowing us to estimate depletion factors of approximatedly 1.3--2.6 for CO and 1.1--3.3 for CS in the 2$\times$10$^{5}$--5$\times$10$^{5}$ years interval, slightly lower than in obsevations. However, as it is apparent from Figure \ref{fig:models_2}, the depletion of CO and CS increases with time. Therefore, it is important to note that the inferred depletion factors are more consistent with the observations at later times. Equivalently, a distribution of dust-grain sizes in the cloud, beyond the assumed grain radius of 0.1\,$\mu$m, may also be invoked to explain the differences between the depletion factors predicted by the models and those inferred from the observations. The depletion of CO and CS correlates very well with the increase in the observed abundance of H$_2$CO, CH$_3$OH and H$_2$CS, due to the efficient hydrogenation of the parent diatomics on ices \citep{Watanabe2002, Cuppen2009, Lamberts2018}, showing the same trend as in the observations where these molecules emission correspond very well with the depletion of CO and CS. 
In the model, grain chemistry is also extremely effective in the formation of HNCO and explain the higher core abundance across all considered times in our simulations, following the preferential observation of this molecule in the core region. The reason for the enhanced formation of HNCO is that, in addition to gas-phase electron recombination:
\begin{align}
    \ce{H2NCO+ + e- &-> HNCO + H }, \\
    \ce{HNCOH+ + e- &-> HNCO + H },
\end{align}
that dominates the formation of this species \citep{Marcelino2009,Marcelino2010}, a new set of reactions on and in grains have a major importance in the production of gas-phase HNCO, namely cosmic ray sputtering from the ice mantle and chemical desorption from the grain reaction:
\begin{equation}
    \ce{N(s) + HCO(s) -> HNCO(g)}. \label{eq:hnco}
\end{equation}
This explains the correlation of HNCO with the CO depletion (HCO is formed from CO hydrogenation) and the preferential presence of this molecule in the core. We note that it is not clear if such an insertion reaction proceeds as described in the network. A detailed investigation of that reaction goes beyond the goals of this work, but it suffices to say that, if that reaction would not contribute to the abundance of HNCO, the ratio $n_{\rm model,env}$ / $n_{\rm model,core}$ should increase. For this reason, the formation of HNCO is also less robust to the selected physical parameters in the model. For instance, the choice of T$_{\rm d}$ of 10 K is made to avoid an excessive slow-down of grain-chemistry routes arising from the use of a single representative binding energy for simple species, or excessive (H$_{2}$) adsorption. However, variations of just a few Kelvin can activate otherwise very slow diffusive processes, like the diffusion of N atoms (Reaction \ref{eq:hnco}). As a result, if the real temperature of TMC-1 is slightly lower than 10 K, the formation of HNCO could also be lower than in our current model. 
It is important to remark that, at all times, reactions 1 and 2 dominate the formation of HNCO, and that these reactions are enhanced at higher densities. This is another indication of the similar trend found for HNCO and cyanopolyynes as indicated in Section \ref{O-bearing}.
The rest of the species considered in this section are more robust to the choice of dust temperature.

A very interesting case is found for the case of cyanopolyynes, where the observations are reproduced at earlier times ($n_{\rm model,env}$ / $n_{\rm model,core}$ < 1.0; presence in the core) but $n_{\rm model,env}$ and $n_{\rm model,core}$ are balanced at approximately 5\,$\times$\,10$^{5}$ yr for both HC$_{3}$N and HC$_{5}$N. The reason for the presence of such a balance point and its absence at earlier times is chemical, and can attributed to a higher importance of certain reactions as a function of density and visual extinction. By looking at the reaction rates at different times for the densities considered in the envelope and core substructures, we conclude that the main formation routes of HC$_3$N (in particular, but applicable to other cyanopolyynes) are the gas-phase reactions (with different importance at different characteristic times)
\begin{align}
    \ce{N + C4H &-> C + HC3N}, \label{eq:1} \\
    \ce{N + C5H &-> C2 + HC3N}, \label{eq:2} \\
    \ce{HC3NH+ + e- &-> H + HC3N}, \label{eq:3}\\
    \ce{CN + C2H2 &-> HC3N + H}, \label{eq:4}\\
    \ce{H + C3N- &-> HC3N + e-}, \label{eq:5}
\end{align}
whereas the main destruction pathways are
\begin{align}
    \ce{C + HC3N &-> C4N + H}, \label{eq:6}\\
    \ce{H+ + HC3N &-> HC3N+ + H}, \label{eq:7} \\
    \ce{H3+ + HC3N &-> HC3NH+ + H2},  \label{eq:8} \\
    \ce{C+ + HC3N &-> C4N+ + H}.  \label{eq:9} 
\end{align}
Interestingly, the list above clearly shows that depletion onto grains is not the main destruction pathway for cyanopolyynes, as gas-phase chemistry proceeds significantly faster. This is the main difference with the case of CO, CS and related molecules. The relative importance of the different formation reactions at the two considered times is simultaneously responsible of the prevalence in the core at 2\,$\times$\,10$^5$ yr and the balance point at 5\,$\times$\,10$^5$ yr. The differences mainly arise from the varying availability of reactants, which can be roughly traced back to the higher abundance of atomic carbon in the envelope at later times due to its lower visual extinction. In contrast to the depletion of CO and CS, for cyanopolyynes we find a better agreement with the observations when adopting an earlier characteristic time.

While the model can reproduce the observational depletion trends for all molecules that we considered in this section, a complete qualitative match is achieved only when we assume different times for different molecules. The prediction of \ce{HC3N} but in general of cyanopolyynes and polyyines like \ce{C4H} and \ce{C4H-} is much more sensitive to the kinetic parameters of the chemical network, affecting the overall timescales on which these molecules should peak in different chemical models. This is in contrast with the depletion of CO and CS, which depend on less reactions and are better constrained,
although certain global modeling parameters, such as grain size distribution might bring our modeling results closer to the observational evidence presented in previous sections.

\section{Conclusions}

The spatial distribution of the molecular emission in TMC-1 depends on the line opacity of the observed species and its chemistry. This effect is particularly important in the emission of the $J$=1-0 and 2-1 lines of
CS (see Sect. \ref{depletion_CS}) and in the hyperfine lines
of the $N$=1-0 transition of CN (see Appendix \ref{low_opacity_CN} and Fig. \ref{fig:cn_vel}). 

We have found that the emission in carbon chains can be explained as due to three filaments at different velocities, the blue ($BF$ at 5.5 \kms), the cyanopolyyne ($CF$ at 5.83 \kms), and the red filament ($RF$, at 6.1 \kms). The three filaments can be interpreted as a single object experiencing a velocity gradient perpendicular to them, most likely caused by gravitational collapse. 
Maps of the molecular emission with angular resolution $\sim$10$''$ and velocity resolution better than 0.1 \kms\, are
needed to clearly distinguish the possible additional structure of these filaments, and to conclude about a physical common origin, or
to a chance aligment of three different clouds. However, the present data clearly point
toward a common physical origin of the three components.

The three filaments are seen in all species we have mapped. Nevertheless, the emission of the O-bearing species CH$_3$OH, H$_2$CO,
CH$_3$CHO, and HOCO$^+$ is concentrated below the cyanopolyyne filament ($CF$) at all velocities. It seems that their emission surround that of cyanopolyynes at
red velocities. However, HNCO and H$_2$CCO are concentrated along the CF filament with a similar behaviour than other carbon chains.

One of the main results of our work is the discovery that carbon monoxide and carbon monosulfide are highly depleted toward the North-West part of the $CF$ filament. This has been shown by a comparison of the emission of the optically thin $J$=1-0 line of C$^{17}$O
and C$^{17}$O, with the emission of cyanopolyynes and that of several optically thin isotopologues. Although difficult to evaluate due to the presence of the different
cloud velocity components, we consider that CO is depleted by a factor $\sim$5-15. The depletion of CS is nicely marked by the lack of emission of C$^{34}$S and
$^{13}$CS toward the nortwest of TMC-1 where CS has is emission maxima. The
interpretation of the emission requires, at least, a two layer model. HCS$^+$ and H$_2$CCS follow exactly the spatial distribution of the rare isotopologues of CS.

Our chemical models show that the depletion of CO and CS can largely be explained by simple adsorption onto dust grains, together with their associated grain-surface chemistry. In contrast, the prevalence of carbon chains in the $CF$ filament seems to originate primarily from chemical effects that regulate their net production rates, with a limited influence of depletion on the grain surface. This fundamental difference in the chemical evolution of the different families of molecules can be used to rationalize the observations, noting that different characteristic times are needed to reconcile all observations. 

\begin{acknowledgements}

This work is based on observations with the Yebes\,40m radio telescope (projects 19A003, 20A014, 20D023, 21A011, 21D005, 22A007, 22B029, and 23A024) and the IRAM\,30m radio telescope. The 40m radio telescope at Yebes Observatory is operated by the Spanish Geographic Institute (IGN, Ministerio de Transportes y Movilidad Sostenible). IRAM is supported by INSU/CNRS (France), MPG (Germany), and IGN (Spain). We thank Spanish Ministerio de Ciencia, Innovaci\'on, y Universidades for funding support through projects PID2019-106110GB-I00, PID2019-106235GB-I00, 
PID2022-137980NB-100, PID2024-156686NB-I00, and PID2023-147545NB-I00. G.M acknowledges grant RYC2022-035442-I funded by MICIU/AEI/10.13039/501100011033 and ESF+ and project 20245AT016 (Proyectos Intramurales CSIC). This work is supported by ERC through grants 610256 (NANOCOSMOS) and 101218790 (ISOCOSMOS) funded by the European Union. Views and opinions expressed are however those of the author(s) only and do not necessarily reflect those of the European Union or the European Research Council Executive Agency. Neither the European Union nor the granting authority can be held responsible for them. 

\end{acknowledgements}

\begin{appendix}
\section{Line identification and line modeling}\label{app:line_identification}

Line identification in this work has been performed using the MADEX code \citep[][see also Sect. \ref{sec:Ancilliary}]{Cernicharo2012}, and the CDMS and JPL catalogues \citep{Muller2005,Pickett1998}. 
MADEX has been developed during the last
40 years \citep{Cernicharo2012} and it contains information on the rotational transitions
and line strengths for 1982 different species with a total of 6818 spectroscopic entries
corresponding to the isotopologues and vibrationally excited states of these
molecules (when laboratory data is available for them). MADEX stores rotational constants that are used
to predict the rotational transitions of the molecules through the corresponding molecular Hamiltonians. It also can diggest standard files 
from other spectroscopic databases (CDMS or JPL). MADEX also contains all available collisional
rate coefficients needed to solve the statistical equilibrium equations when estimating
emerging intensities of rotational lines in molecular clouds. It
has been used to identify lines, and to model their observed profiles and intensities, in this work. 

To model the molecular emission of CO, its isotopologues, CH$_3$OH, cyanopolyynes, cations and anions, and to derive radial velocities and line widths for the different velocity components (see Table \ref{CO_columndensities}), line parameters have been derived by fitting a Gaussian line profile to the data using the GILDAS package. A velocity range of $\pm$20\,\kms\, around each feature was considered for the fit of a polynomial  baseline. Negative features produced in the folding of the frequency switching data were blanked before baseline removal. Previous determinations of the velocity of the cloud derived from QUIJOTE data have been reported by \citet{Cernicharo2020a}.

The intensity scale used in this study is $T_A^*$. Consequently, the telescope parameters and source properties have been used when modeling the emission of the different species to produce synthetic spectra in this temperature scale. The source has been assumed in previous works to be circular with a uniform brightness temperature. The adopted radius is 40$''$ \citep{Fosse2001}. In this work we validate this assumption from a detailed study of the spatial distribution of several molecules (see Sect.\,\ref{Maps}). The observed line intensities have been modeled using a local thermodynamical equilibrium hypothesis (LTE), or a large velocity gradient approach (LVG) when collisional rates are available. In the later case, MADEX uses the formalism for molecular emission described by \citet{Goldreich1974}.

\section{Line profiles of other studied species}\label{app:line_profiles}

The two main emission components seen in CO isotopologues, the BF and RF at 5.7 and 6.25 \kms, are also observed in other molecules. This is the case, for example, of the methanol lines studied here (see Appendix \ref{app:methanol}), in good agreement with the previous results of other methanol transitions by \citet{Soma2015,Soma2018}. The observed lines of SO, $^{34}$SO, and S$^{18}$O also show line profiles consistent with these two velocity components (see Fig.\,\ref{fig:ANC_SPEC} and \citealt{Lique2006}). However, all the remaining lines in Fig.\,\ref{fig:ANC_SPEC} peak at $\sim$5.83 \kms. 
These lines correspond to transitions of DCO$^+$, H$^{13}$CO$^+$, HC$^{18}$O$^+$, $c$-C$_3$H$_2$, $l$-H$_2$CCC, HC$_3$N, and HCC$^{13}$CN. 
The same velocity has been found for all the lines of the cyanopolyynes and their isotopologues \citep{Cernicharo2020a,Cernicharo2023,Cernicharo2024b,Tercero2024}. Maps and line intensities of C$_6$H$^-$, C$_6$H, NNH$^+$ and 
HCCCN have been recently analysed by \citet{Cernicharo2026b}. For HCN, HCO$^+$ and their isotopologues a study of its spatial distribution, and of radiative transfer effects in the
emission of their rotational lines, has been presented by \citet{Marcelino2026}.
Hence, it appears that the high density gas traced by all these high-dipole moment C-bearing molecules has a subtle different behavior, and a different spatial distribution (see Sect.\,\ref{Maps}), than that of the isotopologues of CO, CH$_3$OH and SO. Multiple velocity components have been seen in many other molecular species \citep{Langer1995,Peng1998,Dickens2001,Fosse2001,Lique2006,Soma2015,Soma2018}, often showing different velocity for their maximum of emission \citep[see the case of the low frequency transitions of C$_4$H, CS, CCS, CCCS, and other species observed by][] {Langer1995,Peng1998,Dickens2001,Fuentetaja2025}. However, the transitions of species involving large Einstein coefficients could suffer from line opacity effects \citep[see, e.g., the case of S-bearing species shown by][]{Fuentetaja2025} that will affect their line profiles. This happens to some of the lines studied in this work (see Appendices \ref{app:two_layer} and \ref{low_opacity_CN}). Hence, each molecular line profile has to be studied in detail, including its spatial distribution, to distinguish between chemical effects and radiative trapping in the cloud. All molecules having abundances larger than $\sim$10$^{-10}$ and high dipole moments ($\mu\ge$1 D) will be affected by radiative trapping effects that will complicate the interpretation of the data, in particular the shape of the line profiles, the determination of the cloud volume densities and molecular column densities (see Appendix \ref{app:two_layer}).

\subsection{Methanol lines profiles}\label{app:methanol}
\begin{figure}
\centering
\includegraphics[width=0.95\columnwidth]{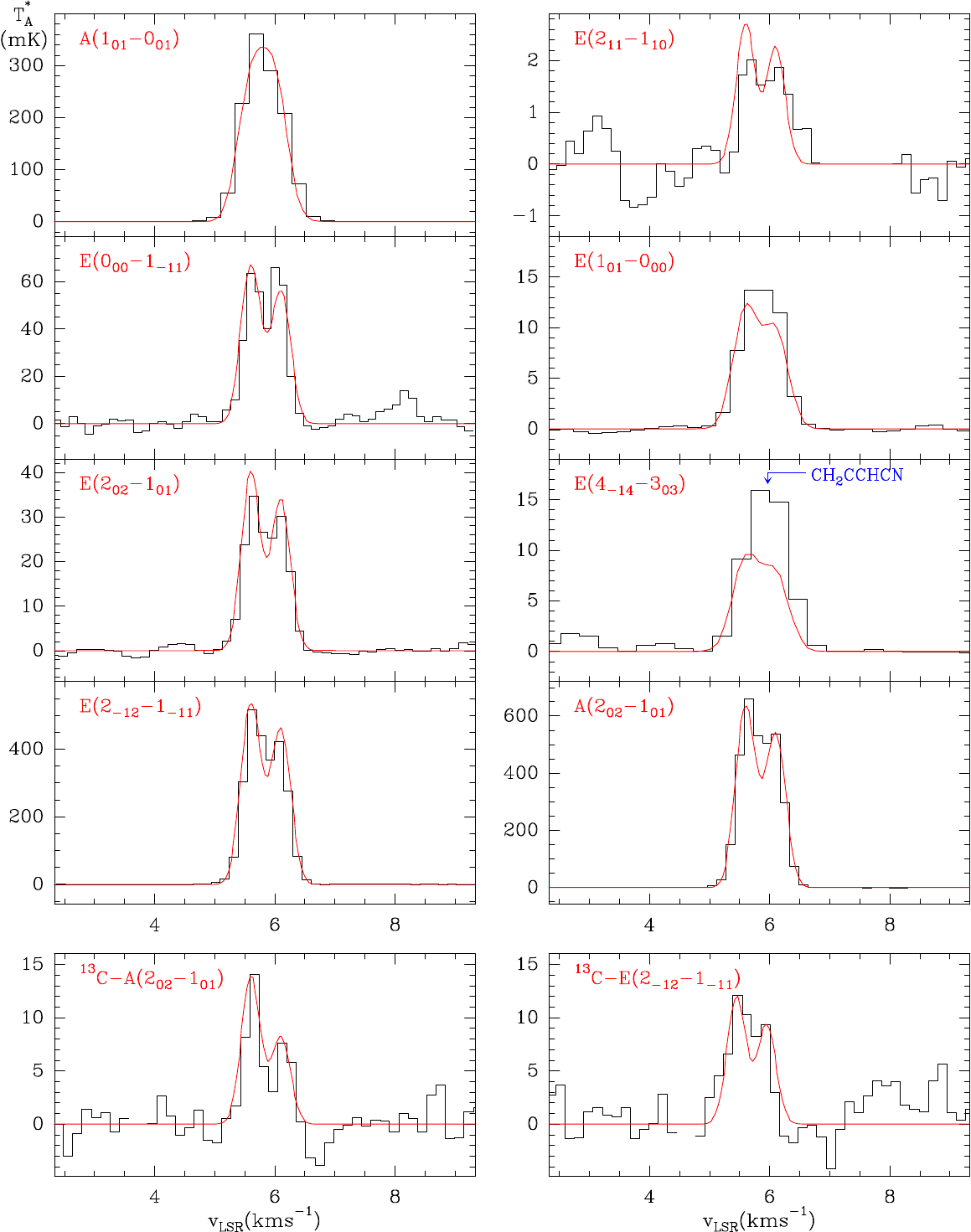} 
\caption{Lines of $A$/$E$-CH$_3$OH observed toward the $CP$  with the Yebes\,40m and IRAM\,30m telescopes. The two bottom panels correspond to the isotopologue $^{13}$CH$_3$OH. The transition $J$, $K$, $p$) and symmetry species is indicated in each panel. The abscissa corresponds to the Local Standard of Rest velocity ($V_{\rm LSR}$). The ordinate is the antenna temperature corrected for atmospheric and telescope losses in mK. Blanked channels correspond to negative features produced in the folding of the frequency switching data. The red curves show the line profiles calculated using LVG and the physical parameters described in Sect.\,\ref{app:methanol}.}
\label{fig:CH3OH_lines}
\end{figure}

The analysis of the line profiles of methanol provides interesting information to understand the spatial distribution of O-bearing species (see Sect.\,\ref{O-bearing}) and to distinguish between gas-phase and ice-phase regulated chemistry in TMC-1. We show in Fig.\,\ref{fig:CH3OH_lines} all the lines of methanol observed toward the $CP$  using Yebes\,40m and IRAM\,30m data. All the lines observed at 3\,mm show a double emission peak at 5.7 and 6.25 \kms, which could result from the different cloud components for methanol or from radiative trapping effects as those shown by several species analysed in this paper. It is important to have a good interpretation of the line profile of methanol to study the maps of its spatial distribution at different velocities (see Sect.\,\ref{O-bearing}). \citet{Soma2015,Soma2018} analysed several lines of methanol which show exactly the same line profile than those observed here at 3\,mm. However, to resolve the origin of the doubly peaked line profile, additional information is required. The detection with a high sensitivity of the two lines of $^{13}$CH$_3$OH shown in Fig.\,\ref{fig:CH3OH_lines} definitively solves the interpretation of the line profile (both were only very marginally detected by \citealt{Soma2015}). We have, similar to the line profile of the isotopologues of CO and other O-bearing species (see Sect.\,\ref{app:line_profiles}), two different velocity components. The best determination from our CH$_3$OH data in Fig.\,\ref{fig:CH3OH_lines} of these velocities is 5.6\,$\pm$\,0.1\,\kms and 6.1\,$\pm$\,0.1\,\kms (they coincide with the blue and red filamentes, $BF$ and $RF$, discussed in Sect.\,\ref{Maps}). 

We can estimate the line opacities, volume densities, and column densities using a LVG approach and the collisional rates of methanol with $p$-H$_2$ \citep{Pottage2004,Rabli2010}. Frequencies and intensities for methanol were taken from the predictions of the CDMS database \citep{Muller2005}. For each velocity component we adopted a linewidth of 0.35 \kms. The best LVG result was obtained for a density $n$(H$_2$)\,=\,1.5\,$\times$\,10$^4$ cm$^{-3}$ for the $BF$ and $RF$ clouds. We found that the $A$ and $E$ species have the same column density in each cloud and that the total column densities of methanol are (2.2\,$\pm$\,0.2)\,$\times$\,10$^{13}$ cm$^{-2}$ and (1.86\,$\pm$\,0.18)\,$\times$\,10$^{13}$ cm$^{-2}$ for the $BF$ and $RF$ clouds, respectively. The opacities derived in our model correspond to the optically thin case for most lines. The strongest lines have opacities of $\sim$ 0.5. The predicted line profiles are shown as a red line in Fig.\,\ref{fig:CH3OH_lines}. The match between observations and model is excellent for the Q-band and 3\,mm lines. 

For $A$-$^{13}$CH$_3$OH the best determination of the column density is (2.7\,$\pm$\,0.5)\,$\times$\,10$^{11}$ cm$^{-2}$ and (1.6\,$\pm$\,0.4)\,$\times$\,10$^{11}$ in the $BF$ and $RF$ clouds, respectively. For $E$-$^{13}$CH$_3$OH the column density in the $BF$ cloud is identical to that of the $A$ species, while it is slightly larger for the $RF$ one, (2.1\,$\pm$\,0.5)\,$\times$\,10$^{11}$ cm$^{-2}$. Hence, the total column densities of $^{13}$CH$_3$OH in the $BF$ and $RF$ clouds is (4.3\,$\pm$\,0.9)\,$\times$\,10$^{11}$ and (4.8\,$\pm$\,0.9)\,$\times$\,10$^{11}$ cm$^{-2}$, respectively, while the $^{12}$C/$^{13}$C ratio is 51 and 39, respectively. The averaged $^{12}$C/$^{13}$C ratio is 45\,$\pm$\,5, in agreement with the previous less accurate values by \cite{Soma2015}. We note that in the study of \citet{Soma2015}, the H$_2$ densities derived from their analysis were different for the blue (1.7$\times$10$^4$ cm$^{-3}$, i.e., very similar to our value of 1.5$\times$10$^4$ cm$^{-3}$) and the red cloud (4.3$\times$10$^4$ cm$^{-3}$, a factor of 3 larger than ours). The difference in the density is probably due to the fact that we observed lines from higher energy levels, which provide a better constraint of the volume density. Their total CH$_3$OH column densities, 2.6$\times$10$^{13}$ and 1.4$\times$10$^{13}$ cm$^{-2}$ for the $BF$ and $RF$ clouds, respectively, are in reasonable good agreement with our results.

\section{A simple two layer model for CS} \label{app:two_layer}

The situation for CS is similar to that analysed by \cite{Cernicharo1987a} for HCO$^+$ and H$^{13}$CO$^+$ and revisited recently, including DCO$^+$, HCN, DCN, and H$^{13}$CN, by \cite{Marcelino2026}. In that study, it was found that the line profiles and intensities reflect a huge opacity in the $J$\,=\,1–0 line of HCN and HCO$^+$. A simple two-layer model was sufficient to explain the observations qualitatively. In the case of CS, a high-density core with $n$(H$_2$)\,=\,2\,$\times$\,10$^4$ cm$^{-3}$ and $N$(CS)\,=\,5.5\,$\times$\,10$^{13}$ cm$^{-2}$, causes the emission of $^{13}$CS and C$^{34}$S \citep{Fuentetaja2025}. This core corresponds in the maps of Fig.\,\ref{fig:ANC_maps} to the narrow filaments $RF$, $CF$, and $BF$. A surrounding envelope with a moderate density of $n$(H$_2$)\,=\,4\,$\times$\,10$^3$ cm$^{-3}$, and a column density of 5\,$\times$\,10$^{13}$ cm$^{-2}$ has still enough molecules of CS (and of HCO$^+$, HCN, and other abundant species; see Appendix\,\ref{low_opacity_CN}) to absorb CS line photons from the core and to re-emit them over a large volume \citep[see the non local radiative transfer models for HCO$^+$ and H$^{13}$CO$^+$ of][]{Cernicharo1987a}. Consequently, the line opacity in the envelope is so high for the abundant CS species that photons emitted in the core never escape the cloud. However, the column density of the second layer (envelope) is not sufficient to produce significant intrinsic emission or notable absorption for $^{13}$CS and C$^{34}$S line photons, which are all coming from the dense filament. The same applies to H$^{13}$CO$^+$, DCO$^+$, DCN, and H$^{13}$CN \citep{Cernicharo1984b,Cernicharo1987a,Gonzalez1993,Marcelino2026}. This situation is typical of any cold cloud with narrow line widths and small velocity gradients. As a consequence, all points of the cloud are, independently of their density, radiatively connected and radiative transfer requires the use of non-local models when line opacities are $\ge$1. The volume density and the column density in the envelope are strongly correlated. Increasing the density to 6\,$\times$\,10$^3$ cm$^{-3}$, then the required column density to obtain a reasonable fit to the lines of CS is 4\,$\times$\,10$^{13}$ cm$^{-2}$.

The two-layer model is an oversimplification of the radiative transfer problem in TMC-1, and a more detailed analysis is required to implement density and velocity gradients from the external parts of the cloud down to the central core (see, e.g. models as those described by  \citealt{Cernicharo1987a} and \citealt{Gonzalez1993}). To achieve these goals, a multi-line study with a higher angular resolution than that used in this work, is needed. Nevertheless, we tested the reliability of the model by comparing the predicted intensities for all the isotopologues of CS with the observations (see \citealt{Fuentetaja2025}). A similar model has been applied to CN (see Appendix\,\ref{low_opacity_CN}) and the isotologues of HCO$^+$ and HCN \citep{Marcelino2026}. We note that density determinations depend on collisional rates and, hence, each molecule could require slightly different densities in the two layers to qualitatively reproduce the observations due to the uncertainties for these collisional rates.

\section{Line opacity and the spatial distribution of CN}\label{low_opacity_CN}
\begin{figure*}
\centering
\includegraphics[width=0.98\textwidth]{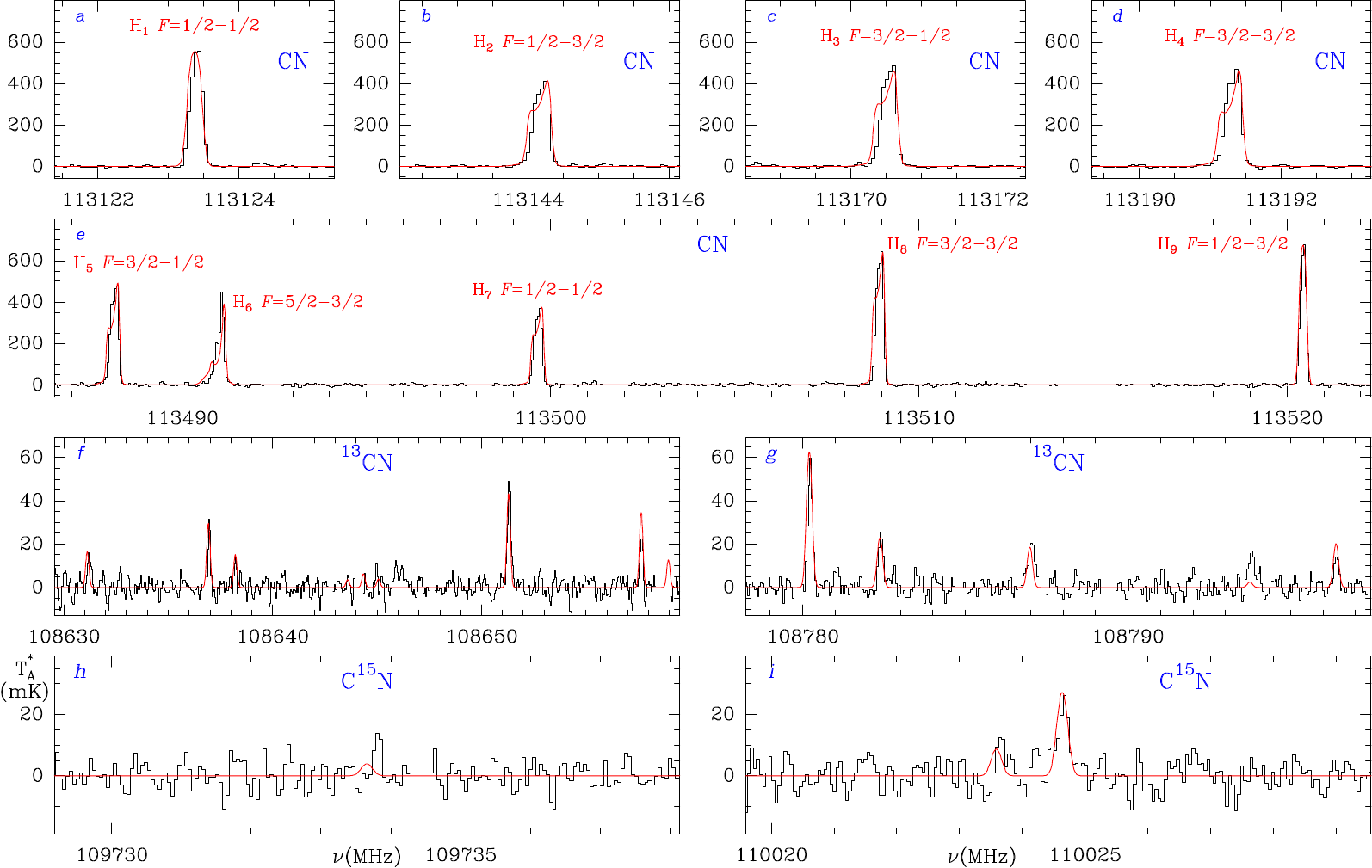}
\caption{Observed fine and hyperfine components of the $N$\,=\,1-0 transition of CN (panels $a,b,c,d$ correspond to the hyperfine components of the $J$\,=\,1/2-1/2 fine transition, while panel $e$ corresponds to the hyperfine components of $J$\,=\,3/2-1/2), $^{13}$CN (panels f and g), and C$^{15}$N (panels h and i). These lines were observed with the IRAM\,30m telescope. For each panel the abscisa is rest frequency in MHz assuming a $V_{\rm LSR}$ for the source of 5.83 \kms \citep{Cernicharo2020a}. The ordinate is antenna temperature in milli Kelvin. The red line is the synthetic spectrum computed under the LVG approximation and the physical parameters given in Sect.\,\ref{low_opacity_CN}.}
\label{fig:cn_spec}
\end{figure*}

From the analysis of CS (see Sect.\,\ref{depletion_CS}), HCO$^+$, and HCN \citep{Marcelino2026}, it is clear that line opacity can introduce important pitfalls in the interpretation of the data. The opacities of the lowest $J$ lines of these species is so large that it is impossible to get any reliable parameter from their observation \citep{Cernicharo1987a,Marcelino2026}. However, for species showing hyperfine structure it is possible in some cases to select the weakest hyperfine component to derive physical and chemical properties from these species (for example HCN, see, e.g., \citealt{Cernicharo1984b,Gonzalez1993,Marcelino2026}. The same can be achieved when the abundance of the molecule is high enough to allow the detection of its rare isotopologues (see, e.g., Fig.\,\ref{fig:ANC_maps}, \ref{fig:ANC_maps_2}, \ref{fig:ANC_maps_3} and Sect.\,\ref{depletion_CS}).

In the following we analyze the spatial distribution, and the line radiative trapping effects, of CN, which have several hyperfine components with different intrinsic line strengths. It is highly recommended to use such kind of molecules in dark clouds where all regions of the cloud are radiatively connected and line opacity can be
huge for the most abundant species.

The CN radical was one of the first molecules detected in space at optical wavelengths in the line of sight of slightly reddened stars \citep{McKellar1940,Adams1941}. It was also one of the first molecules detected in the interstellar medium in the early times of millimeter radio astronomy \citep{Jefferts1970}. \citet{Hily-Blant2008} observed CN and $^{13}$CN toward the prestellar cores L1544 and L183 and concluded that, similarly to NNH$^+$ and NH$_3$, CN was not depleted toward the high-density cores of cold dark clouds. Nevertheless, their analysis indicated that the $N$\,=\,1-0 line was optically thick. Observations of CN and C$^{15}$N have been used to the derive the $^{14}$N/$^{15}$N abundance ratio as a function of the galactocentric distance \citep{Adande2012}. Moreover, the radical CN has been also detected toward C- and O-rich evolved stars and it is one of the most abundant species in the case of C-rich AGBs \citep{Bachiller1997}. In both, interstellar and circunstellar media, the CN radical plays a key role in the growth of cyanopolyynes \citep{Agundez2017,Cernicharo2022b}.

The $N$\,=\,1-0 transition of this molecule has nine hyperfine components and some of them have very weak line intensities which require sensititive data to be detected. Such a large number of hyperfine transitions  in a  very abundant molecule is a real advantage for the interpretation of the data. The analysis could help in distinguishing physical/chemical effects from radiative trapping controlling the emerging line profiles from the cloud. Hence, a study of the spatial distribution of all these hyperfine components could provide us with a clear view of how line opacity limits our interpretation of the chemistry at play in TMC-1.

\begin{figure*}
\centering
\includegraphics[width=0.98\textwidth]{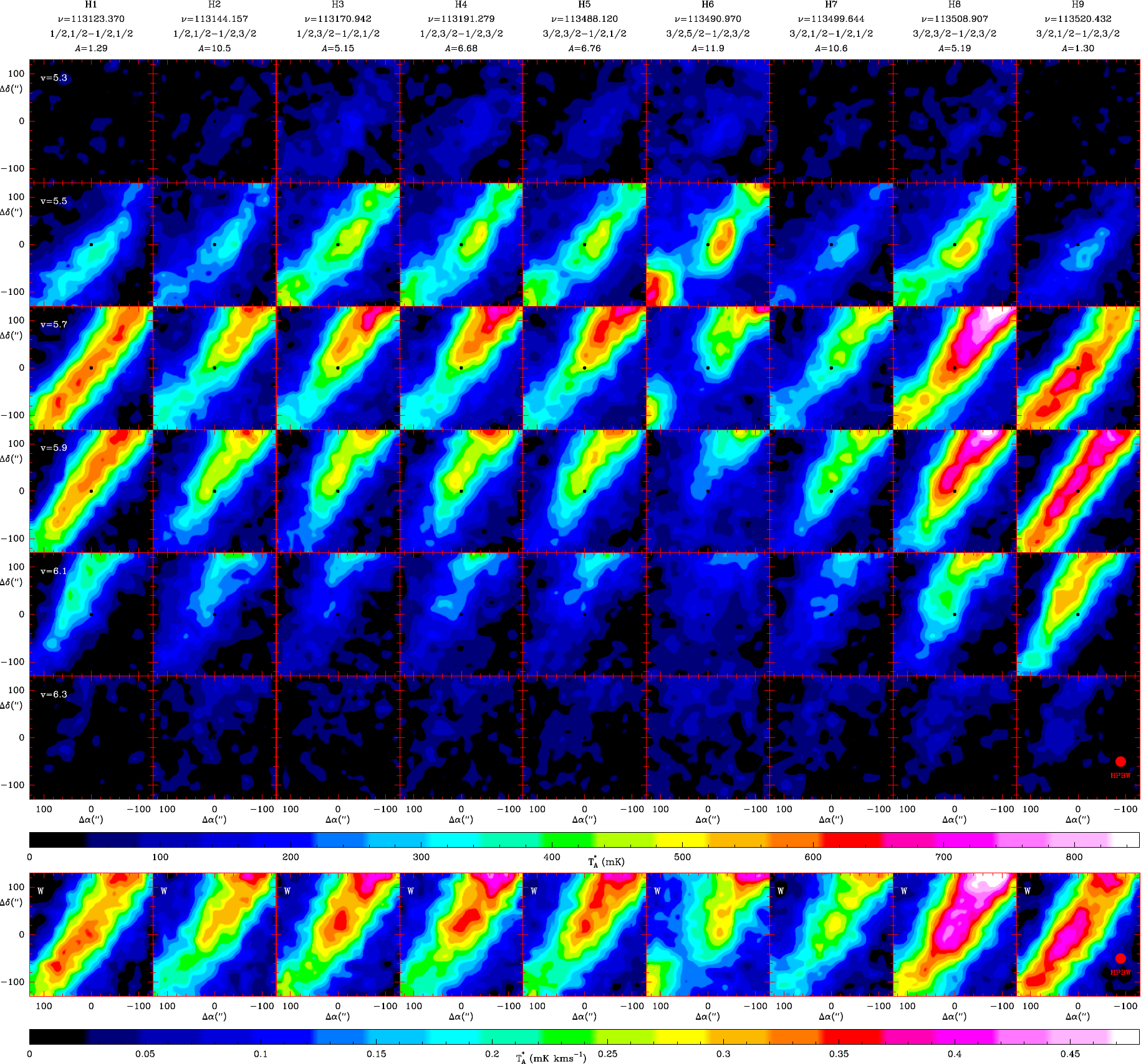}
\caption{Spatial distribution of the nine hyperfine components of the $N$\,=\,1-0 transition of CN as a function of the velocity. Each column corresponds to one of these components while each row corresponds to a different velocity from 5.3 to 6.3 \kms\, by steps of 0.2 \kms. For each column the frequency of the corresponding hyperfine component, its fine and hyperfine quantum numbers, and its Einstein coefficient $A$ (in units of 10$^{-6}$ s$^{-1}$) are given at the top of the panels of the first row. The half power beam width is indicated in the bottom-right panel. The color scale is the same for all panels and is indicated in the wedge below the 6.3 \kms velocity row. Finally, the bottom panels represent the velocity-integrated intensity of each hyperfine component. The scale is the same for all the panels in units of mK \kms.}
\label{fig:cn_vel}
\end{figure*}

Frequency and intensity predictions for CN and its isotopologues $^{13}$CN and C$^{15}$N are from the CDMS \citep{Muller2005}. These predictions arise from a global fit to all rotational lines measured in the laboratory for the three isotopologues  \citep{Dixon1977,Skatrud1983,Johnson1984,Klisch1995,Bogey1984,
Bogey1986,Saleck1994}. We implemented these species in the MADEX code \citep{Cernicharo2012}. A dipole moment of 1.45\,D has been adopted for CN and its isotopologues \citep{Thompson1968}. Collisional rates between CN and $p$-H$_2$ are available for the three isotopologues, CN \citep{Kalugina2012}, $^{13}$CN, and C$^{15}$N \citep{Flower2015} and, consequently, rotational excitation can be computed under the LVG approximation \citep{Goldreich1974}.

The $N$\,=\,1-0 fine and hyperfine components of CN, $^{13}$CN and C$^{15}$N are shown in Fig.\,\ref{fig:cn_spec}. They were observed with the IRAM\,30m telescope. The nine hyperfine components of CN are all well resolved in frequency and are labelled in Fig.\,\ref{fig:cn_spec} as H$_1$ up to H$_9$. The observed line profiles of the CN hyperfine components suggest clear radiative trapping problems. It is a similar case to that discussed CS and isotopologues (see Sect. \ref{depletion_CS} and Appendix \ref{app:two_layer}), HCO$^+$ and isotopologues \citep{Cernicharo1987a,Marcelino2026}, or HCN and isotopologues \citep{Cernicharo1984b,Gonzalez1993,Marcelino2026}. 

In order to interpret the CN data toward the $CP$  position we adopted a similar model to that of CS, i.e., we considered a dense core with $T_k$\,=\,9 K, $n$(H$_2$)\,=\,1.5\,$\times$\,10$^4$ cm$^{-3}$, $N$(CN)\,=\,(9.0\,$\pm$\,0.8)\,$\times$\,10$^{14}$ cm$^{-2}$, $N$($^{13}$CN)\,=\,(12.5\,$\pm$\,1.2)\,$\times$\,10$^{12}$ cm$^{-2}$, and $N$(C$^{15}$N)\,=\,(5.0\,$\pm$\,0.5)\,$\times$\,10$^{12}$ cm$^{-2}$. For the dense core we adopted a line width of $\Delta v$\,=\,0.4 \kms\, for all isotopologues. A second layer with a density of 6\,$\times$\,10$^3$ cm$^{-3}$ and $\Delta v$\,=\,1.1\,\kms was considered. Like for NNH$^+$ this second layer is shifted in velocity by +0.2 \kms which could correspond to gas falling toward the dense core. This model  reproduces rather well the hyperfine lines H$_1$ and H$_9$ of CN, and those of $^{13}$CN and C$^{15}$N (see red lines in Fig.\,\ref{fig:cn_spec}). With the second cloud component we obtain a qualitative good agreement between the observations and the model. This second layer has not a significant effect on $^{13}$CN and C$^{15}$N. However, without the second less dense layer, the hyperfine lines H$_2$ up to H$_8$ of CN will be considerably overestimated. We note that under optically thin conditions LVG calculations for all volume densities provide the same excitation temperatures for all hyperfine components of CN, $^{13}$CN, and the fine structure lines of C$^{15}$N. However, for optically thick conditions the excitation temperatures are significantly different for the components having weak and strong line intensities. In this context, changing in our model the density of the dense layer to 2\,$\times$\,10$^4$ cm$^{-3}$ produces similar results if column densities of CN, $^{13}$CN, and C$^{15}$N are decreased by a factor of $\sim$20\,\%. From our oversimplified model, and considering only the high density component, we obtain a CN/$^{13}$CN and CN/C$^{15}$N abundance ratios of 72\,$\pm$\,13 and 180\,$\pm$\,36, respectively. If we consider that our simple model reproduces rather well the optically thin hyperfine lines H$_1$ and H$_9$, these ratios could be considered as reliable. However, some caution has to be introduced if we focus on a model aiming to reproduce all hyperfine lines of CN as the uncertainty in the density structure of the cloud could affect the emerging line profiles and intensities.

The spatial distribution for velocities between 5.3 and 6.3\,\kms\ for the nine hyperfine components of CN is shown in Fig.\,\ref{fig:cn_vel}. The panels of each column correspond to the spatial distribution at different velocities of one of the hyperfine components. Each row shows the spatial distribution at a given velocity for each of the nine hyperfine components. Frequencies, quantum numbers ($J,F$), and Einstein coefficients are given at the top of the panels of the first row. The spatial distribution of the optically thin components H$_1$ and H$_9$ (lowest Einstein coefficients) is very different to that of the optically thick ones (H$_2$, H$_6$ and H$_7$ are the thickest). While H$_1$ and H$_9$ follows the emission of NNH$^+$, DCCCN, and HCC$^{13}$CN shown in Fig.\,\ref{fig:ANC_maps_2}, and that of HC$_3$N $J$\,=\,9-8 and $J$\,=\,5-4 shown in Fig.\,\ref{fig:ANC_maps}, the other hyperfine components, in particular the strongest H$_5$, H$_6$ and H$_7$, show a spatial distribution that is rather similar to that of CS (see Fig.\,\ref{fig:ANC_maps_2}), HCO$^+$, and HCN  \citep{Marcelino2026}. The effect is event more prominent on the maps of the velocity-integrated emission for each hyperfine transition shown in the bottom panels of Fig.\,\ref{fig:cn_vel}.

Taking into account that the hyperfine components are well resolved in velocity and that no blends are produced in the low-$N$ lines, the observed differences in their spatial distribution is a pure radiative trapping effect across the cloud. The weakest hyperfine lines are the ones transporting information about the densest regions. Detailed non-local radiative transfer calculations have been performed for HCO$^+$ \citep{Cernicharo1987a} and HCN \citep{Gonzalez1993}, which demonstrates that the density structure of the cloud has an important effect in the extent of the optically thick lines as photons arising from the dense core are absorbed and re-emitted by the external layers which have a larger spatial extent. This is what we see in the emission of the H$_6$ and H$_7$ components at 5.7 and 5.9 \kms\, which follows very closely that of the optically thick emission of CS $J$\,=\,2-1, and that of the $J$\,=\,1-0 lines of HCO$^+$ and HCN analysed in detail by \citet{Marcelino2026}. Similarly to the anomalies of HCN \citep{Cernicharo1984b,Gonzalez1993,Marcelino2026}, the transitions with the lowest Einstein coefficients (i.e., the lowest intrinsic intensities $S_{ul}$) are those that emerge with the largest antenna temperature intensities. CN is a very nice example of how careful the analysis of molecular emission has to be done when lines are optically thick and radiative trapping effects dominate the formation of the emergent line profiles.

\end{appendix}

\end{document}